\documentclass[10pt]{IEEEtran}

\usepackage{amsmath,amssymb,amsfonts}
\usepackage{graphicx}
\usepackage{textcomp}

\usepackage[belowskip=-10pt,aboveskip=10pt,small]{caption}
\usepackage{subcaption}
\usepackage{latexsym}
\usepackage{threeparttable}
\usepackage{threeparttablex}
\usepackage{mathtools}
\usepackage{slashbox}
\usepackage{verbatim}
\usepackage{subfloat}
\usepackage{tabu}
\usepackage{booktabs}
\usepackage{float}
\usepackage{url}
\usepackage{multirow}
\usepackage{graphicx}
\usepackage{cite}
\usepackage{footnote}
\usepackage{array}
\usepackage[table]{xcolor}
\usepackage{tabularx}
\usepackage{array}
\usepackage{slashbox}

\usepackage{algpseudocode}
\usepackage{algorithm}

\newcolumntype{P}[1]{>{\centering\arraybackslash}p{#1}}

\usepackage{mathtools, nccmath}

\DeclareMathOperator*{\argmax}{argmax} 
\usepackage{epstopdf}					
\usepackage{epstopdf}					
\usepackage{amsthm} 
\theoremstyle{theorem}
\newtheorem{corollary}{Corollary} 
\newtheorem{lemma}{Lemma} 
\allowdisplaybreaks

\title{{Design and Detection of Unitary Constellations in Non-Coherent SIMO Systems for Short Packet Communications}
\thanks{Son T. Duong and Ebrahim Bedeer are with the Department of Electrical and Computer Engineering, University of Saskatchewan, Saskatoon, Canada S7N 5A9. Emails: \{wve765, e.bedeer\}@usask.ca.}
\thanks{Ha H. Nguyen, deceased, was with the Department of Electrical and Computer Engineering, University of Saskatchewan, Saskatoon, Canada S7N 5A9.}
\thanks{R. Barton is with Cisco Systems Inc. Email: robbarto@cisco.com.}
\thanks{This work was supported by NSERC/Cisco Industrial Research Chair in Low-Power Wireless Access for Sensor Networks.}
}

\author{Son T. Duong, Ha H. Nguyen, Ebrahim Bedeer, and Robert Barton}

\IEEEaftertitletext{\vspace{-2.5\baselineskip}}
\begin{document}

\maketitle

\begin{abstract}
	
This paper proposes a novel design of multi-symbol unitary constellation for non-coherent single-input multiple-output (SIMO) communications over block Rayleigh fading channels.
To facilitate the design and the detection of large unitary constellations at reduced complexity, the proposed constellations are constructed as the Cartesian product of independent amplitude and phase-shift-keying (PSK) vectors, and hence, can be iteratively detected. The amplitude vector is detected by exhaustive search, whose complexity is sufficiently low in short packet transmission scenarios. To detect the PSK vector, we use the posterior probability as a reliability criterion in the sorted decision-feedback differential detection (sort-DFDD), which results in near-optimal error performance for PSK symbols with equal modulation orders. This detector is called posteriori-based-reliability-sort-DFDD (PR-sort-DFDD) and has polynomial complexity. We also propose an improved detector called improved-PR-sort-DFDD to detect a more generalized PSK structure, i.e., PSK symbols with unequal modulation orders. This detector also approaches the optimal error performance with polynomial complexity.
Simulation results show the merits of our proposed multi-symbol unitary constellation when compared to competing low-complexity unitary constellations. 
\end{abstract}

\begin{IEEEkeywords}
Constellation design, decision-feedback differential detection (DFDD), non-coherent detection, short packet communications. 
\end{IEEEkeywords}

\section{Introduction}
{Ultra-reliable low-latency communication (URLLC) is one of the services of the 5G New Radio (5G NR) to support the newly emerging applications such as telesurgery, intelligent transportation, and industrial automation \cite{chen2018ultra}. These emerging applications require a block error rate (BLER) from $10^{-5}$ to $10^{-9}$ and a latency from 1 ms to 10 ms for the transmission of short messages, i.e.,  several tens of bytes. 
Such requirements pose a challenge in designing efficient communication systems with limited power and bandwidth resources since latency and reliability are usually conflicting objectives.}

To meet the stringent latency requirements of URLLC, a variety of technologies, which cover different layers of communication networks, have been proposed and standardized for 5G \cite{li20185g}. At the physical layer, low latency is achieved with the transmission of short packets. However, since short packet transmission does not support powerful channel coding, exploiting different diversity resources to provide high reliability is needed for URLLC. Among different diversity schemes, the use of multiple antennas in the form of single-input multiple-output (SIMO) or multiple-input multiple-output (MIMO) systems with coherent detection has been extensively investigated in recent years \cite{nasir2021cell,ostman2021urllc,ren2020joint}. 
Many Internet of Things (IoT) transmission scenarios can be seen as SIMO systems as end devices are equipped with a single antenna while a typical base station or a gateway can be equipped with multiple antennas.
The performance of coherent detection of multiple-antenna systems is strongly affected by the accuracy of the estimated channel state information (CSI) \cite{zhang2014pilot}. In order to obtain accurate instantaneous CSI at the receiver, the transmitter has to send a sequence of training symbols or pilots, whose length can be comparable to the number of symbols in the short packet.
Hence, using a long sequence of training symbols to estimate the CSI in short packet communications (SPCs) decreases the spectral efficiency (SE).

{Non-coherent communication schemes, which do not require pilot symbols for channel estimation, can reduce the SE loss caused by using the pilot symbols in SPCs \cite{xu2019sixty}.
Among different non-coherent communications, energy-based encoding and decoding have been well-investigated in massive SIMO systems for {fast fading} channels where the coherence time is less than the duration of one information symbol \cite{manolakos2016energy,chowdhury2016scaling,jing2016design,xie2020non}. In particular, the channel hardening property resulting from using a large number of receiving antennas is exploited to decode the signal. However, non-coherent energy-based schemes suffer from loss of the SE since only the amplitude of the signal carries information. For wireless channels having coherence time covering multiple symbol durations, non-coherent multi-symbol encoding and decoding have been investigated to transmit additional information in the phase, thus improving the SE. In differential phase-shift keying (DPSK), the information is encoded into the phase difference between two consecutive symbols, and differential detection is carried out to estimate the phase difference at the receiver \cite{baeza2017analysis}. Another widely investigated multi-symbol encoding and decoding scheme in the literature is unitary constellations. In unitary constellations, all the transmitted multiple-symbol signals have the same energy level, and the received signal can be decoded by using a non-coherent multi-symbol maximum likelihood (ML) detector \cite{hochwald2000unitary}. Multi-level unitary constellations can be considered to be generalized schemes of unitary constellations, where information is encoded into constellations of energy levels and unitary constellations. The design of energy levels has been only investigated in \cite{borran2003design,duong2023multi}, however, both papers do not study the design of unitary constellations. 
}

In this paper, we focus on the design and detection of unitary constellations. The investigation of unitary constellations is justified as it is proven to be the optimal non-coherent transmission scheme in high SNR regimes \cite{hochwald2000unitary,zheng2002communication}. Different designs of unitary constellations have been proposed in the literature, and they can be classified into two main groups: structured unitary constellations and unstructured unitary constellations. Unstructured unitary constellations are typically obtained by numerically solving optimization problems of minimizing pairwise error probability (PEP). {For MIMO communications, different design criteria have been investigated such as high-SNR asymptotic expression of PEP \cite{cuevas2023union}, the Chernoff bound of PEP \cite{wu2008unitary}, singular values of two signal's correlation \cite{hochwald2000unitary, hochwald2000systematic}, and chordal distance \cite{hochwald2000systematic}. Please note in the SIMO system, the design criteria based on singular values in \cite{hochwald2000unitary, hochwald2000systematic} are simplified to the chordal distance.}
While the unstructured unitary constellations can achieve optimal PEP performance, they have exponential complexity in design and therefore are only feasible for small constellation sizes. On the other hand, structured unitary constellations have a pre-determined structure mainly to reduce the complexity of the design process, thus allowing the design of larger unitary constellations \cite{hochwald2000systematic,attiah2016systematic,zhang2011full,kammoun2007non,kim2010unitary,ngo2019cube,li2019design}. As expected, structured unitary constellations have inferior PEP performance when compared to unstructured unitary constellations. Additionally, some structures of unitary constellations are restricted to some specific cases of blocklength and constellation size. For example, the structure in \cite{li2019design} is restricted to a blocklength of only two symbols, while the structures in \cite{kim2010unitary} and \cite{ngo2019cube} are restricted to a blocklength of power of two.

The optimal ML detector of unitary constellations is complex for both unstructured and structured unitary constellations with large constellation sizes \cite{ngo2019cube}.
Therefore, various approaches have been proposed to reduce the detection complexity of the unitary constellations \cite{gohary2009noncoherent, li2019design,kammoun2007non,ngo2019cube}. Quasi-ML detection was proposed in \cite{gohary2009noncoherent} to reduce the complexity of solving the ML detection. However, the quasi-ML detection requires the storage of all possible constellation points, which makes it still impractical for large constellations. Another approach is to design the structure of unitary constellations in a way that allows lower complexity in detection (in addition to low complexity in design). Please note that this approach is not applicable to unstructured unitary constellations and many structured unitary constellations, such as, the Fourier-based method \cite{hochwald2000systematic}, geometric motion \cite{attiah2016systematic}, and the co-prime PSK constellations \cite{zhang2011full}. 
In \cite{li2019design}, a unitary constellation based on phase-shift keying (PSK) was proposed that reduces the complexity of the ML detector into the complexity of detecting PSK symbols.
In \cite{kammoun2007non} and \cite{ngo2019cube}, exponential mapping and cube-split mapping were proposed to map QAM symbols into a non-coherent unitary sequence, respectively, thus reducing the complexity of encoding and decoding the unitary constellation to the complexity of encoding and decoding QAM symbols. 

A special case of structured unitary constellations is a sequence of PSK symbols.
The non-coherent detection of this structure can be classified into two major groups: (i) multi-symbol differential detection (MSDD) \cite{ho1992error,divsalar1994maximum,papailiopoulos2013maximum} and (ii) decision-feedback differential detection (DFDD) \cite{adachi1995decision,adachi1996decision,schober1999decision}. The MSDD algorithm is equivalent to the ML detector and therefore has optimal error performance at the cost of exponential complexity. 
{In the DFDD algorithm, the decisions of detected PSK symbols are used as feedback to detect the next PSK symbol in sequential order. The DFDD algorithm has low complexity but suffers from performance loss due to the error propagation because if a PSK symbol is detected incorrectly, this incorrect decision is still used as feedback to detect the next PSK symbol. The sort-DFDD algorithms in \cite{schenk2013noncoherent, fischer2014noncoherent} resolve the error propagation issue by detecting the most reliable PSK symbols first and then using them as feedback. Because reliable symbols are less likely to cause errors, the sort-DFDD algorithm reduces the effect of error propagation. {The reliability criterion was defined as the quantization error in \cite{schenk2013noncoherent} and the log-likelihood ratio (LLR) in \cite{fischer2014noncoherent}. Please note that the sort-DFDD algorithm in \cite{schenk2013noncoherent} was only designed for PSK sequence with equal modulation orders and equal amplitude; while the sort-DFDD algorithm in \cite{fischer2014noncoherent} was further restricted to 4-PSK constellations. }} 


{
In this paper\footnote{In \cite{duong2023_design_unitary}, we presented only the design of the proposed unitary constellation. In the current paper, we additionally present low-complexity detectors of the proposed unitary constellation.}, we propose a novel multi-symbol unitary constellation with low complexity in both the design and detection in the non-coherent SIMO system for SPCs.
The main contributions of this paper are summarized as follows:
\begin{enumerate}
	\item We propose a novel unitary constellation based on two design rules. In design rule 1, we confine the unitary constellations into the pairwise product of amplitude vector and phase vector, and the amplitude vector and the phase vector are independent of each other, i.e., they belong to independent constellation sets. In design rule 2, we restrict the symbols of the phase vector to belong to a set of PSK constellations, which can have either equal modulation orders or unequal modulation orders.
    \item We formulate an optimization problem to maximize the minimum chordal distance (MCD) between any two vectors of the proposed unitary constellations. We exploit the proposed structure, imposed by the two design rules, to solve the unitary constellation design problem at reduced computation complexity when compared to general unstructured unitary constellations in \cite{hochwald2000unitary}.
    \item We exploit the first design rule and propose a novel detector, named, iterative unitary amplitude-phase detector (IUAP), to decouple the joint detection of the amplitude and phase vectors of the ML detector into independent amplitude vector detection and phase vector detection.
    \item For the phase vector detection, we exploit the second design rule and propose two modified versions of the sort-DFDD algorithm. Inspired by \cite{fischer2014noncoherent}, we consider the posterior probability as the reliability criterion and propose the PR-sort-DFDD algorithm when the phase vector has equal modulation orders. For the case that the phase vector has unequal modulation orders, we use the information from undetected PSK symbols with low modulation order to assist with the detection of PSK symbols with high modulation orders. This is what we call the proposed improved-PR-sort-DFDD algorithm. 
    \item Simulation results show that our proposed unitary constellations have a larger MCD than other low-complexity unitary constellations from the literature. Our proposed detectors achieve near error performance of the optimal ML detector while having lower complexity. Additionally, our proposed unitary constellation with the proposed detectors achieves a better error performance when compared to the low-complexity unitary constellations and pilot-based QAM and PSK schemes.
\end{enumerate}
}

\emph{Notation:} Matrices, column vectors, and scalar variables are denoted by uppercase bold letters (e.g., $\mathbf{A}$), lowercase bold letters (e.g., $\mathbf{a}$), and lowercase letters (e.g., $a$), respectively. The notations $(\cdot)^*$, $(\cdot)^T$, and $(\cdot)^H$ represent the conjugate, transpose, and conjugate transpose, respectively. We use $|\cdot|$, $\Vert \cdot \Vert$, $\det(\cdot)$, $\rm{tr}(\cdot)$, and $\circ$ to denote absolute value, Euclidean norm, determinant, trace, and inner product, respectively. The angle of a complex scalar variable $a$ is denoted as $\angle a$. We use $\mathbb{R}_{+}^{m \times n}$ and $\mathbb{C}^{m \times n}$ to indicate the set of non-negative real matrices and complex matrices with dimension $m \times n$, respectively. The matrix $\mathbf{I}_K$ denotes the $K \times K$ identity matrix, while $\mathbf{1}_K$ and $\mathbf{0}_K$ denote all ones $K$-dimensional vector and all zeros $K$-dimensional vector, respectively. The cardinality of the constellation set $\Omega$ is represented as ${\rm{card}}\{\Omega\}$, while the exclusion of an element $b$ from the set $\Omega$ and the exclusion of a subset $\tilde{\Omega}$ from the set $\Omega$ is denoted as $\Omega \backslash b$ and $\Omega \backslash \tilde{\Omega}$, respectively. The Gamma and signum functions are denoted as $\Gamma(.)$ and $\text{sgn}(.)$, respectively.
The probability density function (PDF) of a random continuous variable is denoted as $f(.)$, while the probability of an event is denoted as $P(.)$. The circularly symmetric complex Gaussian distribution with mean $\mu$ and variance $\sigma^2$ is represented as $\mathcal{C N}(\mu,\sigma^2)$, and we use $\mathbb{E}(\cdot)$ to denote the expectation.

The remainder of the paper is organized as follows. The system model and overview of unitary constellations are presented in Section~\ref{section:system_model}. We propose our structure of the proposed unitary constellations and its design method in Section~\ref{section:unitary_design}. The detection of the proposed unitary constellation is presented in Section~\ref{section:detection}. Simulation results and related discussion are given in Section~\ref{section:simulation_result}. Finally, Section~\ref{section:conclusion} concludes the paper.

\section{System Model}
\label{section:system_model}

We consider the uplink transmission of a SIMO system in which a single antenna transmitter communicates with an $M$-antenna base station. Let us denote $\mathbf{h} \in \mathbb{C}^{M \times 1}$ as the vector representing the channel coefficients between the transmitter and the receiver. We assume that the channel is unknown to both of the transmitter and receiver and distributed as $\mathbf{h} \sim \mathcal{C N}(\mathbf{0}_M, \mathbf{I}_M)$. We additionally assume block fading where the channel coefficients remain constant over a block of consecutive $K$ symbols, then change to an independent realization in the coming block of consecutive $K$ symbols.
Let $\mathbf{v} \in \mathbb{C}^{K \times 1}$ be a sequence of $K$ transmit symbols drawn from a unitary constellation set $\Omega_v$ that satisfies the constraint $\Vert \mathbf{v} \Vert^2 = 1$. The received signal $\mathbf{Y} \in \mathbb{C}^{M \times K}$ can be written as:
\begin{equation}
    \mathbf{Y} = \mathbf{h} \mathbf{v}^T + \mathbf{N},
    \label{eqn:channel_model}
\end{equation}
where $\mathbf{N} \in \mathbb{C}^{M \times K}$ is the matrix of the additive noise. The elements of $\mathbf{N}$ are independent zero-mean circular Gaussian random variables with variance $\sigma^2$. In the following two subsections, we provide the unitary constellation design and detection criteria, respectively.

\vspace{-10pt}
{\subsection{Unitary Constellation Design Criterion: Minimum Chordal Distance (MCD)}}
Hochwald and Marzetta in \cite{hochwald2000unitary} proved that the Chernoff bound of the PEP between two unitary signals $\mathbf{v}_a$ and $\mathbf{v}_b$ when using the ML detector is given by:
\begin{align}
    P_e (\mathbf{v}_a \xrightarrow{} \mathbf{v}_b) \leq \frac{1}{2} \left[ 1+\frac{D^2_v(\mathbf{v}_a,\mathbf{v}_b)}{4\sigma^2(1+\sigma^2)} \right]^{-M},
    \label{eqn:Chernoff_bound}
\end{align}
where $D_v(\mathbf{v}_a,\mathbf{v}_b)$ is the chordal distance between $\mathbf{v}_a$ and $\mathbf{v}_b$:
\begin{align}
    D_v(\mathbf{v}_a,\mathbf{v}_b) = \sqrt{ 1 - \left| \mathbf{v}_a^H \mathbf{v}_b \right|^2 }.
    \label{eqn:squared_chordal_distance}
\end{align}
One can see that the PEP between $\mathbf{v}_a$ and $\mathbf{v}_b$ increases as $D_v(\mathbf{v}_a,\mathbf{v}_b)$ decreases. Hence, the unitary constellation $\Omega_v$ can be designed by maximizing the minimum chordal distance (MCD) between all vectors $\mathbf{v}$ in the unitary constellation $\Omega_v$, which is given by:
\begin{subequations}
    \label{eqn:unitary_set_problem_transformed0}
    \begin{align}
        \hat{\Omega}_v & = \argmax_{\Omega_v} \left\{ \min_{\substack{\mathbf{v}_a \neq \mathbf{v}_b \\ \mathbf{v}_a,\mathbf{v}_b \in \Omega_v}} D_v(\mathbf{v}_a,\mathbf{v}_b) \right\}, \\
        \text{s.t.} & \ \Vert \mathbf{v} \Vert = 1, \ \forall \mathbf{v} \in \Omega_v, \ {{\rm{card}}} \{\Omega_v\} = 2^{l_v}, \label{eqn:unitary_set_problem_transformed0_constraint2}
    \end{align}
\end{subequations}
where $l_v$ is the number of bits allocated to $\Omega_v$. The problem in (\ref{eqn:unitary_set_problem_transformed0}) requires the optimization of $2^{l_v}$ constellation points; hence, its complexity grows exponentially with $l_v$.

{
By directly solving (\ref{eqn:unitary_set_problem_transformed0}) using numerical techniques, we obtain the general unitary constellation \cite{hochwald2000unitary}. The general unitary constellation is unstructured because there are no constraints in (\ref{eqn:unitary_set_problem_transformed0}) that confine the constellations into a specific unitary structure. 
In other words, adding any constraints to reduce the design or detection complexity results in what is called structured unitary constellations. 
While the general unitary constellation outperforms structured unitary constellations in terms of the MCD and error performance, its exponential complexity in design and detection makes it impractical for many applications. Thus, designing structured unitary constellations that can achieve good MCD and error probability performance with low complexity in design and/or detection, when compared to the general unitary constellations, is crucial. Such low complexity design of the proposed structured unitary constellation is what will be discussed in Section \ref{section:unitary_design}.
}

{\subsection{Unitary Constellation Detection Criterion: Maximum Likelihood (ML) Detection}}
For the general case of $\mathbf{v}$, the probability density function (PDF) of the received signal $\mathbf{Y}$ given the sequence of the $K$ transmit symbols $\mathbf{v}$ can be written as follows \cite{hochwald2000unitary}:
\begin{align}
    f(\mathbf{Y}|\mathbf{v}) 
    = \frac{\exp\left(-\frac{\text{tr}(\mathbf{Y}^H\mathbf{Y})}{\sigma^2} + \frac{\text{tr}\left( \mathbf{Y}^H \mathbf{Y} \ \mathbf{v}^* \mathbf{v}^T \right)}{\sigma^2(\sigma^2 + \Vert \mathbf{v} \Vert^2)} \right)}{\pi^{KM}\left[ (\sigma^2 + \Vert \mathbf{v} \Vert^2)\sigma^{2K-2} \right]^{M}},
    \label{eqn:density_func}
\end{align}
and the optimal ML detector of $\mathbf{v}$ is given by \cite{hochwald2000unitary}:
\begin{equation}
    \mathbf{v}_{\text{ML}} = \argmax_{\mathbf{v} \in \Omega_v} \left( \frac{\text{tr} \left( \mathbf{Y}^H \mathbf{Y} \ \mathbf{v}^* \mathbf{v}^T \right)}{\sigma^2(\sigma^2 + \Vert \mathbf{v} \Vert^2 )} - M \text{ln} (\sigma^2 + \Vert \mathbf{v} \Vert^2 ) \right).
    \label{eqn:ML_detector}
\end{equation}
As mentioned earlier, we focus on the design and detection of the unitary constellation $\Omega_v$, where $\Vert \mathbf{v} \Vert=1,  \forall \mathbf{v} \in \Omega_v$; hence, the {optimal ML detector} in (\ref{eqn:ML_detector}) can be simplified as follows:
\begin{equation}
    \mathbf{v}_{\text{ML}} = \argmax_{\mathbf{v} \in \Omega_v} \ \text{tr} \left( \mathbf{Y}^H \mathbf{Y} \ \mathbf{v}^* \mathbf{v}^T \right) = \argmax_{\mathbf{v} \in \Omega_v} \Vert \mathbf{Y}^H \mathbf{v} \Vert^2.
    \label{eqn:GLRT_detector}
\end{equation}
{The low complexity  detection of the proposed structured unitary constellation is what will be discussed in Section \ref{section:detection}.}

\section{Design of the Proposed Unitary Constellation}
\label{section:unitary_design}
In this section, we propose a novel design of unitary constellations.

\vspace{-10pt}
\subsection{Proposed Unitary Constellation Design}
One can represent a vector $\mathbf{v}$ of $K$ symbols as a pair-wise product of amplitude vector $\mathbf{u}$ and phase vector $\mathbf{p}$ as:
\begin{align}\label{eq:car}
    \mathbf{v} = \mathbf{u} \circ \mathbf{p},
\end{align}
where $\mathbf{u} = \left[u_0, \dots, u_{K-1}\right]^T \in \mathbb{R}_{+}^{K \times 1} $ and $\mathbf{p} = \left[e^{j\phi_0}, \dots, e^{j\phi_{K-1}}\right]^T \in \mathbb{C}^{K \times 1}$. The two design rules of our proposed unitary constellations $\Omega_v$
are mainly to reduce the construction and detection complexity as will be shown in Sections \ref{section:unitary_design} and \ref{section:detection}, respectively, and they can be explained as follows:

\begin{itemize}
\item {\textit{Design rule 1:} We assume that the amplitude vector $\mathbf{u}$ and phase vector $\mathbf{p}$ are independent of each other, i.e., they belong to independent constellation sets. Hence, the proposed unitary constellation $\Omega_v$ with the first design rule can be formally written as:
\begin{align}
\label{eqn:proposed_structure}
    \Omega_v = \left\{ \mathbf{v} \ | \ \mathbf{v} = \mathbf{u} \circ \mathbf{p}, \mathbf{u} \in \Omega_u, \mathbf{p} \in \Omega_p \right\},
\end{align}
where $\Omega_u=\{\mathbf{u}_{0},\dots,\mathbf{u}_{2^{l_u}-1}\}$ is the set of amplitude constellations and $\Omega_p = \{\mathbf{p}_{0},\dots,\mathbf{p}_{2^{l_p}-1}\}$ is the set of phase constellations, with $l_u$ and $l_p$ being the number of bits allocated to $\Omega_u$ and $\Omega_p$, respectively. Please note that $l_u + l_p = l_v$.}

\item {\textit{Design rule 2:} We confine $\mathbf{p}$ such that its symbols belong to a set of PSK constellations defined as follows:
\begin{align} \label{eq:phases}
    \mathbf{p} = [e^{j\phi_0}, \dots, e^{j\phi_{K-1}}]^{\text{T}}, \phi_k \in \Omega_{\phi_k},
\end{align}
where $\Omega_{\phi_k}$ is the set of $2^{l_{\phi_k}}$ PSK constellation for the $k$-th symbol, $k = 0, ..., K-1$, of the phase vector $\mathbf{p}$. Clearly, $l_{\phi_k}$ has to satisfy $l_p = \sum_{k=0}^{K-1} l_{\phi_k}$. 
}
\end{itemize}

\begin{figure}[t]
	\centering
	\includegraphics[height=4cm, width=6.5cm]{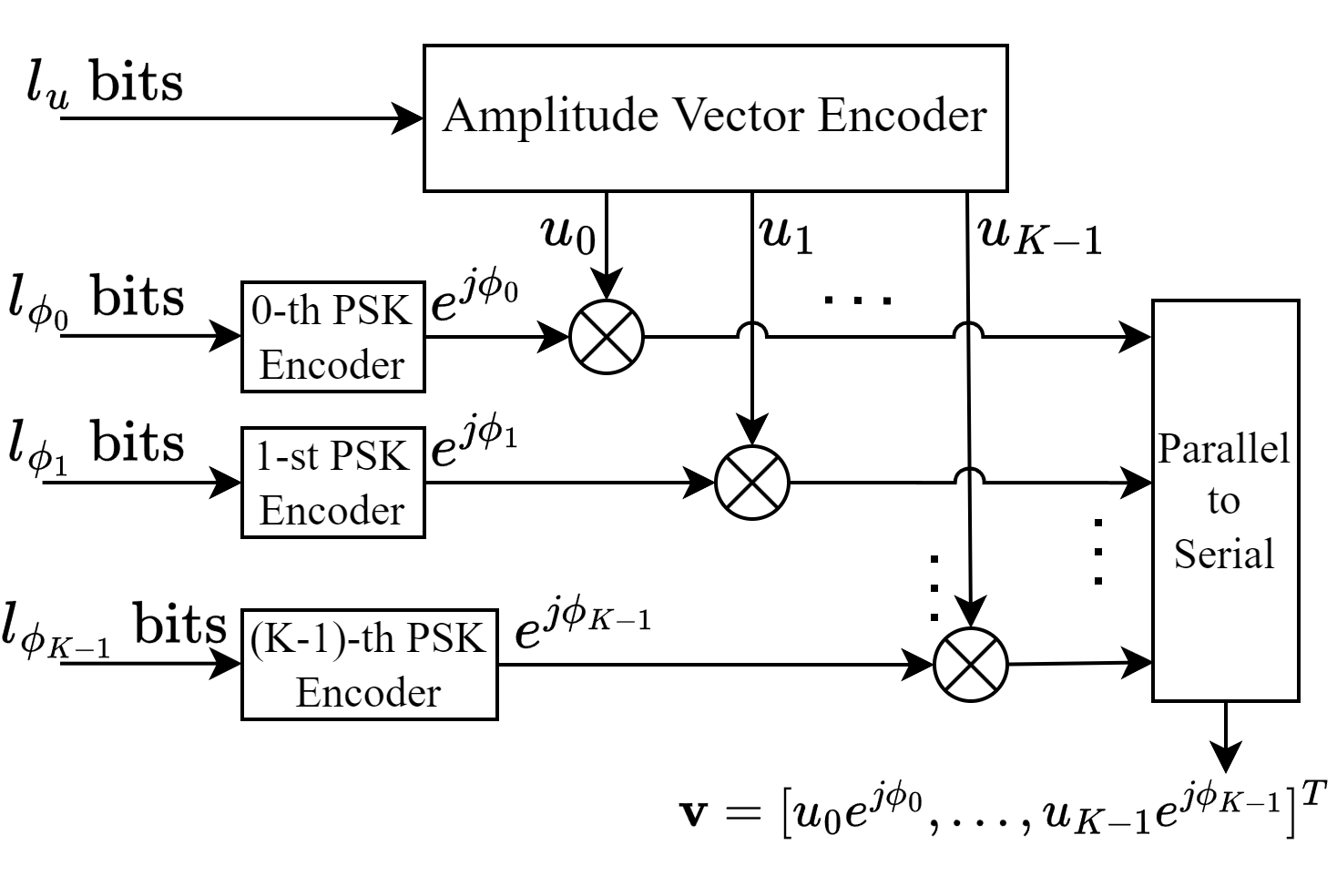}
	\vspace{-15pt}
 \caption{Encoder for the proposed unitary constellations.} 
	\label{fig:Unitary_APSK_encoder}
\end{figure}

\begin{figure}[t]
    \begin{subfigure}{0.5\textwidth}
        \centering
        \includegraphics[width=8cm]{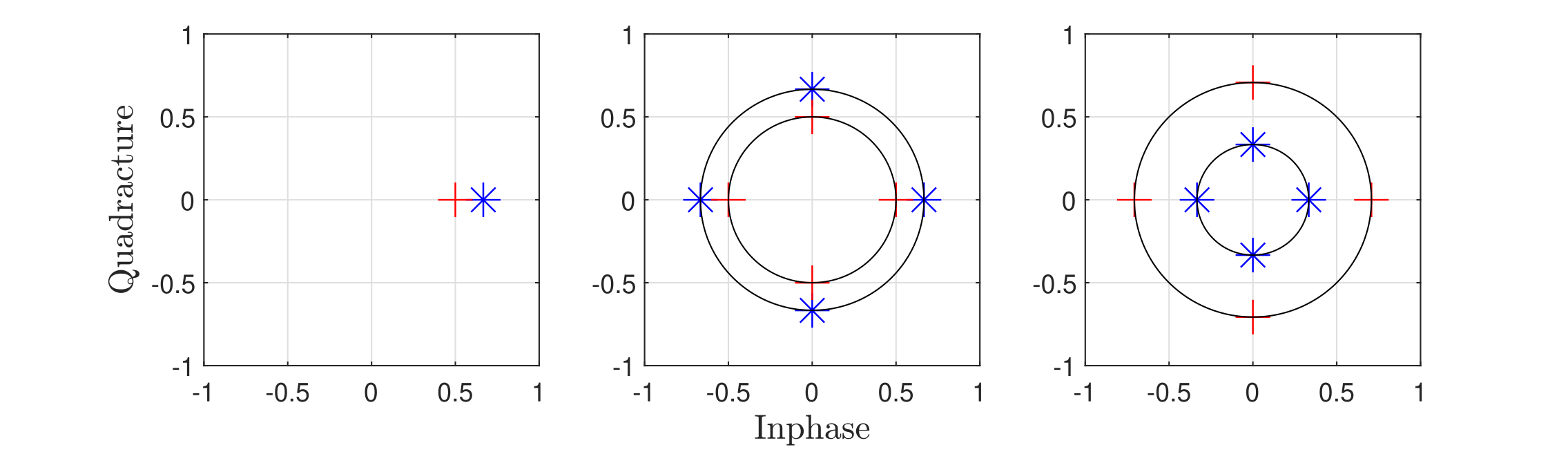}
        \caption{Proposed design with \textbf{\emph{equal modulation orders}} ($l_{\phi_1} = l_{\phi_2} = 2$).}
        \vspace{10pt}
        \label{fig:illustration_4PSK_4PSK}
    \end{subfigure}
    
    \begin{subfigure}{0.5\textwidth}
        \centering
        \includegraphics[width=8cm]{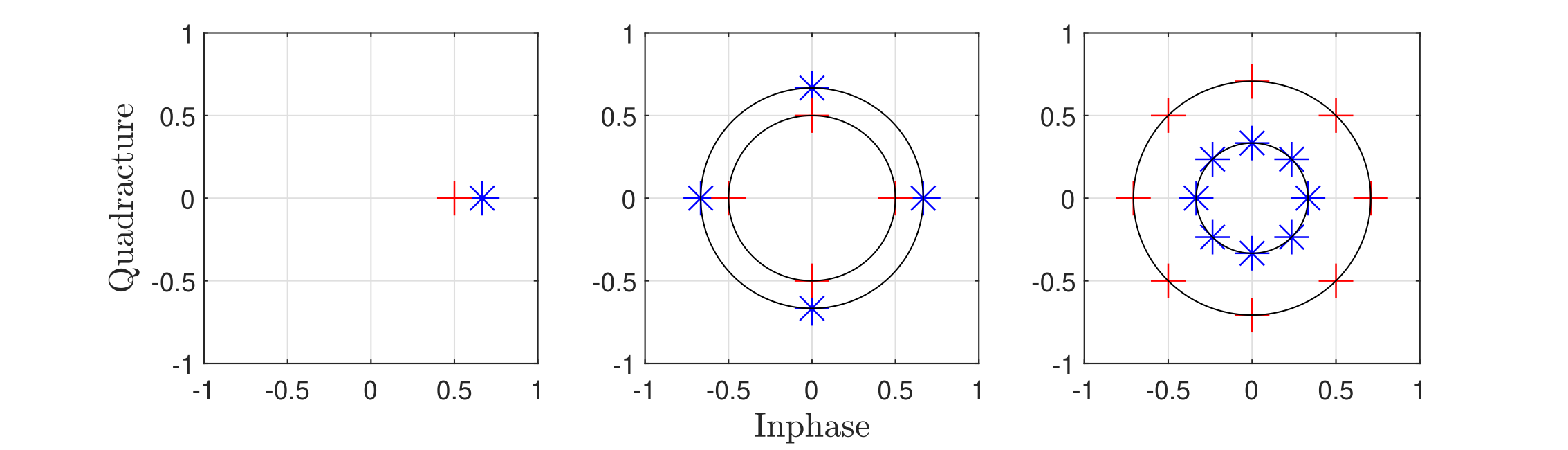}
        \caption{Proposed design with \textbf{\emph{unequal modulation orders}} ($l_{\phi_1}=2$, $l_{\phi_2}=3$).}
        \label{fig:illustration_4PSK_8PSK}
    \end{subfigure}
        \caption{Example of our proposed unitary constellations with $l_u=1$. \textcolor{blue}{Blue}: constellation points with $\mathbf{u}_0=[\frac{2}{3},\frac{2}{3},\frac{1}{3}]$. \textcolor{red}{Red}: constellation points with $\mathbf{u}_1=[\frac{1}{2},\frac{1}{2},\frac{\sqrt{2}}{2}]$.}
    \label{fig:illustration_PSK}
\end{figure}

Fig. \ref{fig:Unitary_APSK_encoder} depicts the encoder of our proposed unitary constellation. Fig. \ref{fig:illustration_4PSK_4PSK} and \ref{fig:illustration_4PSK_8PSK} provide examples of three-symbol constellation ($K = 3$) with \textbf{\emph{equal modulation orders}} ($l_{\phi_1} = l_{\phi_2} = 2$) and \textbf{\emph{unequal modulation orders}} ($l_{\phi_1}=2$ and $l_{\phi_2}=3$), respectively. In both examples, $l_u=1$ corresponds to 1 bit encoded into the amplitude, while the remaining bits are used for encoding the PSK symbols in the phase vector $\mathbf{p}$. Please note that the first symbol in this three-symbol unitary constellation has a reference known phase, i.e., $l_{\phi_0}=0$.

\subsection{MCD Optimization of the Proposed Unitary Constellation}
\label{subsection:constructing_constellation}
{In this subsection, we formulate the MCD optimization problem of the proposed unitary constellation $\Omega_v$ under fixed bit allocation by considering the two design rules formally defined in \eqref{eqn:proposed_structure} and \eqref{eq:phases}.} Let $\mathbf{l}_{\phi} = [l_{\phi_0},\dots,l_{\phi_{K-1}}]$ be a vector indicating the number of bits encoded into the $K$ PSK symbols and $\Omega_p(\mathbf{l}_{\phi})$ be the set of the PSK constellations of the phase vector $\mathbf{p}$ constructed from $\mathbf{l}_{\phi}$.
Clearly, $\Omega_p(\mathbf{l}_{\phi})$ is determined by knowing $\mathbf{l}_{\phi}$. The set of amplitude constellation $\Omega_u$ is unknown and has a size of $2^{l_u}$. Hence, to construct the proposed unitary constellation $\Omega_v$, we need to optimize $\Omega_u$ for given $\mathbf{l}_{\phi}$ and $l_u$.
Without loss of generality, we assume that $l_{\phi_0} \leq \dots \leq l_{\phi_{K-1}}$, and since rotating all symbols of $\mathbf{v}$ by a constant phase does not affect the MCD in \eqref{eqn:squared_chordal_distance} or the  ML detector in (\ref{eqn:GLRT_detector}), we set the $0$-th symbol phase as $\phi_0=0$, i.e., $l_{\phi_0}=0$.

Under the design rules defined in (\ref{eqn:proposed_structure}) and \eqref{eq:phases}, we reformulate the constellation design problem in \eqref{eqn:unitary_set_problem_transformed0} as follows:
\begin{subequations}
    \label{eqn:optProb_fixedBitAllocation}
    \begin{align}
        & \Omega_u{(l_u,\mathbf{l}_{\phi})} = \argmax_{\Omega_u} \left\{ \min_{ \substack{ (\mathbf{u}_a,\mathbf{p}_a) \neq (\mathbf{u}_b,\mathbf{p}_b) \\ \mathbf{u}_a,\mathbf{u}_b \in \Omega_u \\ \mathbf{p}_a,\mathbf{p}_b \in \Omega_p{(\mathbf{l}_{\phi})} } } D_{v}(\mathbf{u}_a\circ\mathbf{p}_a, \mathbf{u}_b\circ\mathbf{p}_b) \right\}, \\
         & \text{s.t.} \ \Omega_u = \left\{ \mathbf{u} | \mathbf{u} \in \mathbb{R}_{+}^{K \times 1}, \Vert \mathbf{u} \Vert = 1 \right\}, \label{eqn:unitary_constraint}\\
         & \qquad {\rm{card}} \{\Omega_u\} = 2^{l_u}. \label{eqn:size_constraint} 
    \end{align}
\end{subequations}
{The problem in (\ref{eqn:optProb_fixedBitAllocation}) only involves the optimization of $2^{l_u}$ constellation points belonged to $\Omega_u$. Thus, the complexity of solving (\ref{eqn:optProb_fixedBitAllocation}) of our proposed design is significantly lower than the complexity of solving (\ref{eqn:unitary_set_problem_transformed0}) of the general design, given that $2^{l_u}$ is chosen to be much smaller than $2^{l_v}$.}

In the following, we analyze the objective function in (\ref{eqn:optProb_fixedBitAllocation}), which is the MCD of all the constellation points. For any two different constellation points, i.e., $(\mathbf{u}_a,\mathbf{p}_a) \neq (\mathbf{u}_b,\mathbf{p}_b)$, the MCD only belongs to one of two possibilities. The first possibility is when the two points have different amplitude vectors, i.e., $\mathbf{u}_a \neq \mathbf{u}_b$, and in this case,  we define $D_u(\mathbf{u}_a,\mathbf{u}_b)$ as the MCD between the two points with the amplitude vectors $\mathbf{u}_a$, $\mathbf{u}_b$ as follows:
\begin{align}
    D_u(\mathbf{u}_a,\mathbf{u}_b) & = \min_{ \mathbf{p}_a,\mathbf{p}_b \in \Omega_p{(\mathbf{l}_{\phi})} } D_v(\mathbf{u}_a \circ \mathbf{p}_a,\mathbf{u}_b \circ \mathbf{p}_b) \nonumber \\
    & = \sqrt{ 1 - \left| \sum_{i=0}^{K-1} u_{i,a} u_{i,b} \right|^2 } = D_v(\mathbf{u}_a ,\mathbf{u}_b).
    \label{eqn:distance_by_amplitude_difference}
\end{align}
where $u_{i,a}$ and $u_{i,b}$ are the $i$-th elements of $\mathbf{u}_a$ and $\mathbf{u}_b$, respectively. As can be seen in \eqref{eqn:distance_by_amplitude_difference}, $D_u(\mathbf{u}_a,\mathbf{u}_b)$ does not depend on the phase vector $\mathbf{p}$.
The second possibility is when the two points have the same amplitude vector but different phase vectors, i.e., $\mathbf{u}_a = \mathbf{u}_b = \mathbf{u}$ and $\mathbf{p}_a \neq \mathbf{p}_b$, and in this case, we define $D_p(\mathbf{u})$ as the MCD among all points with the same amplitude vector $\mathbf{u}$ as follows:
\begin{align}
    D_p(\mathbf{u}) & = \min_{ \substack{\mathbf{p}_a \neq \mathbf{p}_b \\ \mathbf{p}_a,\mathbf{p}_b \in \Omega_p{(\mathbf{l}_{\phi})} } } D_{v} \left( \mathbf{u} \circ \mathbf{p}_a,\mathbf{u} \circ \mathbf{p}_b \right) \nonumber \\
    & = \min_{ \substack{\mathbf{p}_a \neq \mathbf{p}_b \\ \mathbf{p}_a,\mathbf{p}_b \in \Omega_p{(\mathbf{l}_{\phi})} } } \sqrt{1 - \left| \sum_{i=0}^{K-1} u_{i}^2 e^{j(\phi_{i,b} - \phi_{i,a})} \right|^2}.
    \label{eqn:distance_by_phase_difference}
\end{align}
A closed-form expression of $D_p(\mathbf{u})$  can be written as
    \begin{align}
        D_p(\mathbf{u}) = \min \left[ \min_{k \in \mathcal{K}} D^{(k)}_{p,1} (\mathbf{u}), \ \min_{l \in \mathcal{L}} D^{(l)}_{p,2} (\mathbf{u})  \right],
        \label{eqn:closed_form_phase_distance}
    \end{align}
    where $\mathcal{K} = \{k_1,k_2,\dots,k_{Q_1}\}$  is a set of the non-zero indices of $\mathbf{l}_{\phi}$, $\mathcal{L} = \{l_1, l_2, \dots, l_{Q_2}\}$ is a set of the non-zeros indices in ascending order of $\mathbf{l}_{\phi}$ without repetition,    
 $D^{(k)}_{p,1} (\mathbf{u})$ is the MCD of the $k$-th PSK symbol and given by:
    \begin{align}
        D^{(k)}_{p,1} (\mathbf{u}) = \sqrt{ 4 u_k^2 \left( 1 - u_k^2 \right) \sin\left(\frac{\pi}{2^{l_{\phi_k}}}\right)^2},
    \label{eqn:first_phase_distance}
    \end{align}
    and $D^{(l)}_{p,2} (\mathbf{u})$ is the MCD of the reference symbols of the $2^l$-PSK symbols and given by:
    \begin{align}
        D^{(l)}_{p,2} (\mathbf{u}) = \sqrt{ 4 \left( \sum_{ \{ i|l_{\phi_i} < l \} } u_i^2 \right) \left( 1 - \sum_{ \{ i|l_{\phi_i} < l \} } u_i^2 \right) \sin\left(\frac{\pi}{2^{l}}\right)^2}.
        \label{eqn:second_phase_distance}
    \end{align}
The proof is provided in Appendix~\ref{appendix_A1} and Appendix~\ref{appendix_AA1}.

We rewrite the proposed unitary constellation design problem in (\ref{eqn:optProb_fixedBitAllocation}) with the help of $D_u(\mathbf{u}_a,\mathbf{u}_b)$ and $D_p(\mathbf{u})$ as follows:
\begin{subequations}  \label{eqn:optProb_fixedBitAllocation3}
    \begin{align}
        & \Omega_u{(l_u,\mathbf{l}_{\phi})} \nonumber \\
        & = \argmax_{\Omega_u} \left\{ \min \left[ \min_{ \substack{ \mathbf{u}_a \neq \mathbf{u}_b \\ \mathbf{u}_a,\mathbf{u}_b \in \Omega_u } } D_{u}(\mathbf{u}_a, \mathbf{u}_b), \min_{ \substack{ \mathbf{u} \in \Omega_u } } D_{p}(\mathbf{u}) \right] \right\}, \\
        & \text{s.t. } (\ref{eqn:unitary_constraint}), (\ref{eqn:size_constraint}). \nonumber 
        \end{align}
\end{subequations}
{To facilitate obtaining the solution of the unitary constellation design problem in (\ref{eqn:optProb_fixedBitAllocation3}), we make the following two reformulations. Firstly, we re-write the constraint $\Vert \mathbf{u} \Vert^2=1$ as $1 - \epsilon_v \leq \Vert \mathbf{u} \Vert \leq 1$, where $\epsilon_v$ is positive number tends to zero. Secondly, we use the epigraph representation \cite{boyd2004convex} to simplify the objective function of the max-min optimization problem in (\ref{eqn:optProb_fixedBitAllocation3}). Thus, we rewrite the problem in (\ref{eqn:optProb_fixedBitAllocation3}) as follows:}
\begin{subequations}
    \label{eqn:OptProb2}
    \begin{align}
        & \Omega_u{(l_u,\mathbf{l}_{\phi})} = \argmax_{\Omega_{u}} \ t, \\
        \text{s.t.} \ &  D^2_{u} \left( \mathbf{u}_a, \mathbf{u}_b \right) \geq t, \ \forall \mathbf{u}_a, \mathbf{u}_b \in \Omega_u \label{eqn:OptProb2_constraint1}, \\
        \ & \left( D^{(k)}_{p,1}(\mathbf{u}) \right)^2 \geq t, \ \forall \mathbf{u} \in \Omega_u \label{eqn:OptProb2_constraint2}, \ k \in \mathcal{K}, \\
        \ & \left( D^{(l)}_{p,2}(\mathbf{u}) \right)^2 \geq t, \ \forall \mathbf{u} \in \Omega_u \label{eqn:OptProb2_constraint3}, \ l \in \mathcal{L}, \\
        \ & \Vert \mathbf{u} \Vert^2 \geq 1 - \epsilon_v, \forall \mathbf{u} \in \Omega_u, \label{eqn:OptProb2_constraint4} \\
        \ & \Vert \mathbf{u} \Vert^2 \leq 1, \forall \mathbf{u} \in \Omega_u, \label{eqn:OptProb2_constraint5} \\
        \ & (\ref{eqn:size_constraint}). \nonumber
    \end{align}
\end{subequations}
Please note that the problem in (\ref{eqn:OptProb2}) is a non-convex optimization problem due to the non-convex constraints (\ref{eqn:OptProb2_constraint1}), (\ref{eqn:OptProb2_constraint2}), (\ref{eqn:OptProb2_constraint3}), and (\ref{eqn:OptProb2_constraint4}). In the following, we address these non-convex constraints as follows. Firstly, one can see that the constraints in (\ref{eqn:OptProb2_constraint2}) or (\ref{eqn:OptProb2_constraint3}) can be rewritten as a quadratic inequality, i.e., $4x(1-x) \geq y$. This inequality can be reformulated as $1-x \leq (\sqrt{1 - y}+1)/2$ and $x \leq (\sqrt{1 - y}+1)/2$.
By setting $x = u_k^2$ and noting that $\sum_{i=0}^{K-1} u_i^2=1$, the constraints (\ref{eqn:OptProb2_constraint2}) are reformulated to following convex constraints:
\begin{align}
    & \sum_{i=0,i\neq k}^{K-1} u_{i}^2 \leq \frac{\sqrt{1-t/{\sin(\frac{\pi}{2^{l_{\phi_k}}})^2}} + 1}{2}, \nonumber \\
    & u_{k}^2 \leq \frac{\sqrt{1-t/{\sin(\frac{\pi}{2^{l_{\phi_k}}})^2}} + 1}{2}, \ \forall \mathbf{u} \in \Omega_u, k \in \mathcal{K}. \label{eqn:OptProb2_convexified_constraint2}
\end{align}
Similarly, by setting $x=\sum_{\{i|l_{\phi_i} < l\}} u_{i}^2$, the constraints (\ref{eqn:OptProb2_constraint3}) are reformulated to following convex constraints:
\begin{align}
    & \sum_{\{i|l_{\phi_i} < l\}} u_{i}^2 \leq \frac{\sqrt{1-t/{\sin(\frac{\pi}{2^{l}})^2}} + 1}{2}, \nonumber \\
    & \sum_{\{i|l_{\phi_i} \geq l\}} u_{i}^2 \leq \frac{\sqrt{1-t/{\sin(\frac{\pi}{2^{l}})^2}} + 1}{2}, \ \forall \mathbf{u} \in \Omega_u, l \in \mathcal{L}.
    \label{eqn:OptProb2_convexified_constraint3}
\end{align}
{Additionally, we employ Successive Convex Approximation (SCA) algorithm, which use the first-order Taylor series to approximate the non-convex constraints in (\ref{eqn:OptProb2_constraint1}) and  (\ref{eqn:OptProb2_constraint4}), respectively, to the following convex constraints:}
\begin{align}
    & D_u^2(\bar{\mathbf{u}}_a,\bar{\mathbf{u}}_b)  + \nabla_{\mathbf{u}_a} D_u^2(\bar{\mathbf{u}}_a) (\mathbf{u}_a-\bar{\mathbf{u}}_a) + \nabla_{\mathbf{u}_b} D_u^2(\bar{\mathbf{u}}_b) (\mathbf{u}_b-\bar{\mathbf{u}}_b) \nonumber \\
    &  \leq 1-t, \ \forall \mathbf{u}_a, \mathbf{u}_b \in \Omega_u, \ \forall \bar{\mathbf{u}}_a, \bar{\mathbf{u}}_b \in \bar{\Omega}_u(l_u,\mathbf{l}_{\phi}), \label{eqn:OptProb2_convexified_constraint4}
\end{align}
\begin{align}
    \Vert \bar{\mathbf{u}} \Vert^2 + 2\bar{\mathbf{u}}^T(\mathbf{u} - \bar{\mathbf{u}}) \geq 1-\epsilon_v, \ \forall \mathbf{u} \in \Omega_u , \ \forall \bar{\mathbf{u}} \in \bar{\Omega}_u(l_u,\mathbf{l}_{\phi}). \label{eqn:OptProb2_convexified_constraint1}
\end{align}
{where $\bar{\Omega}_{u}(l_u,\mathbf{l}_{\phi})$ is the solution obtained from the previous iteration of the SCA.} Finally, we approximate (\ref{eqn:optProb_fixedBitAllocation3}) into the convex optimization problem as follows:
\begin{subequations}
    \label{eqn:OptProb2_convexified}
    \begin{align}
        & \tilde{\Omega}_{u}(l_u,\mathbf{l}_{\phi}) = \argmax_{\Omega_{u}} \ t, \\
        \text{s.t. } & \text{(\ref{eqn:OptProb2_constraint5}), (\ref{eqn:size_constraint})}, (\ref{eqn:OptProb2_convexified_constraint2}), (\ref{eqn:OptProb2_convexified_constraint3}), (\ref{eqn:OptProb2_convexified_constraint4}), (\ref{eqn:OptProb2_convexified_constraint1}).
    \end{align}
\end{subequations}
{Since, the problem in (\ref{eqn:OptProb2_convexified}) is convex, it can be solved by standard optimization toolboxes such as the CVX toolbox \cite{grant2014cvx}.} Once $\tilde{\Omega}_{u}(l_u,\mathbf{l}_{\phi})$ of (\ref{eqn:OptProb2_convexified}) is obtained, we replace $\bar{\Omega}_{u}(l_u,\mathbf{l}_{\phi})$ with $\tilde{\Omega}_{u}(l_u,\mathbf{l}_{\phi})$ and repeat the process to iteratively approach a solution of the original non-convex design problem in (\ref{eqn:OptProb2}).

\subsection{Searching Good Bit Allocation}
\label{subsection:bit_allocation}
In this subsection, we find bit allocations that result in unitary constellations with the highest MCD. To find the optimal bit allocations, one can perform an exhaustive search over all possible bit allocations. In other words, for each possible bit allocation, one needs to solve the optimization problem in \eqref{eqn:OptProb2_convexified}, and then choose the bit allocation with the highest MCD. However, such an exhaustive search is not practical because the number of combinations of bit allocations that satisfy $l_u+\sum_{k=0}^{K-1} l_{\phi_k}=l_v$ is very large for typical values $K$ and $l_v$ ($K\geq 3$ and $l_v$ is tens of bits).

To reduce the search space, we propose an approach that eliminates bit allocations with low MCDs. This is achieved by finding and testing an upper bound of the MCD
	 $D_{\text{upper}}(l_u,\mathbf{l}_{\phi})$ which is given as follows:
\begin{align}\label{eq:bound}
    & D_{\text{upper}}(l_u,\mathbf{l}_{\phi}) \nonumber \\
    &= \min \left[ \left( \frac{\pi^{\frac{1}{2}} \Gamma(\frac{K+1}{2})}{\Gamma(\frac{K}{2}) } \right)^{\frac{1}{K-1}} 2^{-\frac{l_u}{K-1}}, \sin \left( \frac{\pi}{2^{l_{\phi,\text{max}}}} \right) \right],
\end{align}
where $\Gamma(.)$ is the Gamma function. The proof of the expression of $D_{\text{upper}}(l_u,\mathbf{l}_{\phi})$  is given in Appendix~\ref{appendix_amplitude_bound}.  If a given bit allocation has a small value of $D_{\text{upper}}$, its corresponding  MCD  also must be small. 
Thus, by testing the upper bound on the MCD in \eqref{eq:bound}, we can eliminate bit allocations that result in lower MCDs, and hence, poor error rate performance without solving the optimization problem in \eqref{eqn:OptProb2_convexified}.
Please note that bit allocations with high values of $D_{\text{upper}}$ are not guaranteed to have high MCD. Hence, the MCD must be obtained by solving \eqref{eqn:OptProb2_convexified} in order to find the bit allocation with the highest MCD.

In Table~\ref{table:optimalBitAllocation}, we provide the bit allocations of $K$ = 3 and 4 and $l_v$ up to 15 only due to space limitations. We generate the best bit allocation up to $K=8$ and $l_v=24$\footnote{The best found bit allocations for other values of $K$ and the codebook of $\Omega_u$ are available at \url{https://github.com/duongthanhson97/UnitaryConstellationsCodebook}.}. {Because of the high complexity and because a high value of $K$ is not necessary for SPC, we do not generate bit allocation for $K>8$.}

\begin{table}[t]
    \caption{Bit allocation with the highest MCD ($K$ = 3 and 4, respectively).}\vspace{-8pt}
    \label{table:optimalBitAllocation}
    \begin{subtable}[h]{\linewidth}
        \centering
        \begin{tabular}{|c|c|c||c|c|c||c|c|c|c}
        \hline
        $l_v$ & $\hat{l}_u$ & $\hat{\mathbf{l}}_{\phi}$ & $l_v$ & $\hat{l}_u$ & $\hat{\mathbf{l}}_{\phi}$ & $l_v$ & $\hat{l}_u$ & $\hat{\mathbf{l}}_{\phi}$ \\
        \hline
        1 & 0 & $0,0,1$ & 5 & 1 & $0,2,2$ & 9 & 3 & $0,3,3$ \\
        \hline
        2 & 0 & $0,1,1$ & 6 & 2 & $0,2,2$ & 10 & 4 & $0,3,3$ \\
        \hline
        3 & 0 & $0,1,2$ & 7 & 2 & $0,2,3$ & 11 & 4 & $0,3,4$ \\
        \hline
        4 & 0 & $0,2,2$ & 8 & 3 & $0,2,3$ & 12 & 5 & $0,3,4$ \\
        \hline
        \end{tabular}
    \end{subtable}
    \hfill 
        \begin{subtable}[h]{\linewidth}
        \centering
        \begin{tabular}{|c|c|c||c|c|c||c|c|c|c}
        \hline
        $l_v$ & $\hat{l}_u$ & $\hat{\mathbf{l}}_{\phi}$ & $l_v$ & $\hat{l}_u$ & $\hat{\mathbf{l}}_{\phi}$ & $l_v$ & $\hat{l}_u$ & $\hat{\mathbf{l}}_{\phi}$ \\
        \hline
        1 & 0 & $0,0,0,1$ & 6 & 0 & $0,2,2,2$ & 11 & 4 & $0,2,2,3$ \\
        \hline
        2 & 0 & $0,0,1,1$ & 7 & 2 & $0,1,2,2$ & 12 & 4 & $0,2,2,3$ \\
        \hline
        3 & 0 & $0,1,1,1$ & 8 & 2 & $0,2,2,2$ & 13 & 4 & $0,3,3,3$ \\
        \hline
        4 & 0 & $0,1,1,2$ & 9 & 3 & $0,2,2,2$ & 14 & 5 & $0,3,3,3$ \\
        \hline
        5 & 0 & $0,1,2,2$ & 10 & 4 & $0,2,2,2$ & 15 & 6 & $0,3,3,3$ \\
        \hline
        \end{tabular}
    \end{subtable}
\end{table}

\section{Detection of the Proposed Unitary Constellations}
\label{section:detection}

In this section, we propose to iteratively detect the amplitude and the phase vectors of the proposed unitary constellation using an iterative unitary amplitude-phase (IUAP) detector. For the detection of the phase vector with equal and unequal modulation orders, we propose two low-complexity detectors, namely, posteriori-based-reliability-sort-DFDD (PR-sort-DFDD) and improved-PR-sort-DFDD, respectively.

\subsection{Iterative Unitary Amplitude-Phase Detection Algorithm}
\label{section:iterative_amplitude_phase_detection}

As mentioned in \eqref{eq:car}, the transmitted signal $\mathbf{v}$ can be represented as the pair-wise product of the amplitude vector $\mathbf{u}$ and the phase vector $\mathbf{p}$. Hence, to jointly find the optimal values of  $\mathbf{u}$ and  $\mathbf{p}$ by directly solving the ML detector in (\ref{eqn:GLRT_detector}) with a complexity order of $\mathcal{O}(2^{l_v})$, i.e., exponential in the number of bits allocated to the vector $\mathbf{v}$, i.e., $l_v$. However, by exploiting the first design rule defined in \eqref{eqn:proposed_structure}, i.e., the amplitude vector $\mathbf{u}$ and the phase vector $\mathbf{p}$ belong to independent constellation sets, we can reduce the complexity of the ML detector.

With the help of \eqref{eqn:proposed_structure}, the {ML detector} in (\ref{eqn:GLRT_detector}) can be rewritten as follows:
\begin{align}
    & \{ \mathbf{u}_{\text{ML}}, \mathbf{p}_{\text{ML}} \} = \argmax_{\mathbf{u} \in \Omega_u,\mathbf{p}\in\Omega_p} \text{tr} \left( \mathbf{Y}^H \mathbf{Y} (\mathbf{u} \circ \mathbf{p})^* (\mathbf{u} \circ \mathbf{p})^T \right) \nonumber \\
    & = \argmax_{\mathbf{u} \in \Omega_u,\mathbf{p}\in\Omega_p} \text{tr} \left( \mathbf{Y}^H \mathbf{Y} \left[ (\mathbf{u} \mathbf{u}^T) \circ (\mathbf{p}^* \mathbf{p}^T) \right] \right),
    \label{eqn:ML_detector_magnitude_phase_split}
\end{align}
which can be equivalently represented as either one of the two following detection problems:
\begin{subequations}
    \begin{align}
        \{ \mathbf{u}_{\text{ML}}, \mathbf{p}_{\text{ML}} \} & = \argmax_{\mathbf{u} \in \Omega_u,\mathbf{p}\in\Omega_p} \text{tr} \left( \left[(\mathbf{Y}^H \mathbf{Y}) \circ (\mathbf{u} \mathbf{u}^T) \right] (\mathbf{p}^* \mathbf{p}^T) \right), \\
        \{ \mathbf{u}_{\text{ML}}, \mathbf{p}_{\text{ML}} \} & = \argmax_{\mathbf{u} \in \Omega_u,\mathbf{p}\in\Omega_p} \text{tr} \left( \left[(\mathbf{Y}^H \mathbf{Y}) \circ (\mathbf{p} \mathbf{p}^H) \right] (\mathbf{u} \mathbf{u}^T) \right).
    \end{align}
\end{subequations}
The idea of the proposed IUAP algorithm is to iteratively find sub-optimal  $\tilde{\mathbf{u}}$ and $\tilde{\mathbf{p}}$ until a certain stopping criterion is satisfied. The sub-optimal $\tilde{\mathbf{u}}$ is found by solving:
\begin{align}
    \tilde{\mathbf{u}} = \argmax_{\mathbf{u} \in \Omega_u} \text{tr} \left( \mathbf{Z}^{(\tilde{\mathbf{p}})} (\mathbf{u} \mathbf{u}^T) \right),
    \label{eqn:u_ML_detector}
\end{align}
where $\mathbf{Z}^{(\tilde{\mathbf{p}})} = \left[(\mathbf{Y}^H \mathbf{Y}) \circ (\tilde{\mathbf{p}} \tilde{\mathbf{p}}^H) \right]$. Similarly, the sub-optimal $\tilde{\mathbf{p}}$ is found by solving:
\begin{align}
    \tilde{\mathbf{p}} = \argmax_{\mathbf{p} \in \Omega_p} \text{tr} \left( \mathbf{Z}^{(\tilde{\mathbf{u}})} (\mathbf{p}^* \mathbf{p}^T) \right),
    \label{eqn:p_ML_detector}
\end{align}
where $\mathbf{Z}^{(\tilde{\mathbf{u}})} = \left[(\mathbf{Y}^H \mathbf{Y}) \circ (\tilde{\mathbf{u}} \tilde{\mathbf{u}}^T) \right]$. In other words, the proposed IUAP solves \eqref{eqn:u_ML_detector} and \eqref{eqn:p_ML_detector} iteratively with a complexity order of  $\mathcal{O}(2^{l_u})$ and $\mathcal{O}(2^{l_p})$, respectively. This represents a total complexity order of $\mathcal{O}(2^{l_u}) + \mathcal{O}(2^{l_p})$ which is much smaller than  the complexity order of $\mathcal{O}(2^{l_v}) = \mathcal{O}(2^{l_u + l_p})$ required to directly solve the ML detection problem in (\ref{eqn:GLRT_detector}).
However, it is shown in Table \ref{table:optimalBitAllocation} that the number of bits  $l_p$ is generally much larger than $l_u$ thus it is necessary to develop a low-complexity phase detector to replace the ML phase detector. This is what will be discussed in the following subsections.

\vspace{-10pt}
{
\subsection{Detection of Phase Vector with Equal Modulation Orders} }

\label{subsection:sorted_DFDD}

In this subsection, we develop the PR-sort-DFDD algorithm to detect the PSK-structured phase vector $\mathbf{p}$ with equal modulation orders. Let us assume that at the beginning of the $n$-th iteration of the PR-sort-DFDD, $n$ PSK symbols have been detected with the indices belong to the set $\bar{\mathcal{D}} = \{{k}_0, \dots, {k}_{n-1} \}$ and their final decisions are $\{ \bar{\phi}_{{k}_0}, ..., \bar{\phi}_{{k}_{n-1}} \}$. The remaining undetected PSK symbols have the indices in the set $\mathcal{D} = \{ d_0, ..., d_{K - n-1} \}$. Then, the PR-sorted-DFDD chooses the symbol with the highest reliability among $\mathcal{D}$ to be detected next.

\begin{figure}[t]
    \centering
    \includegraphics[width=2.5cm]{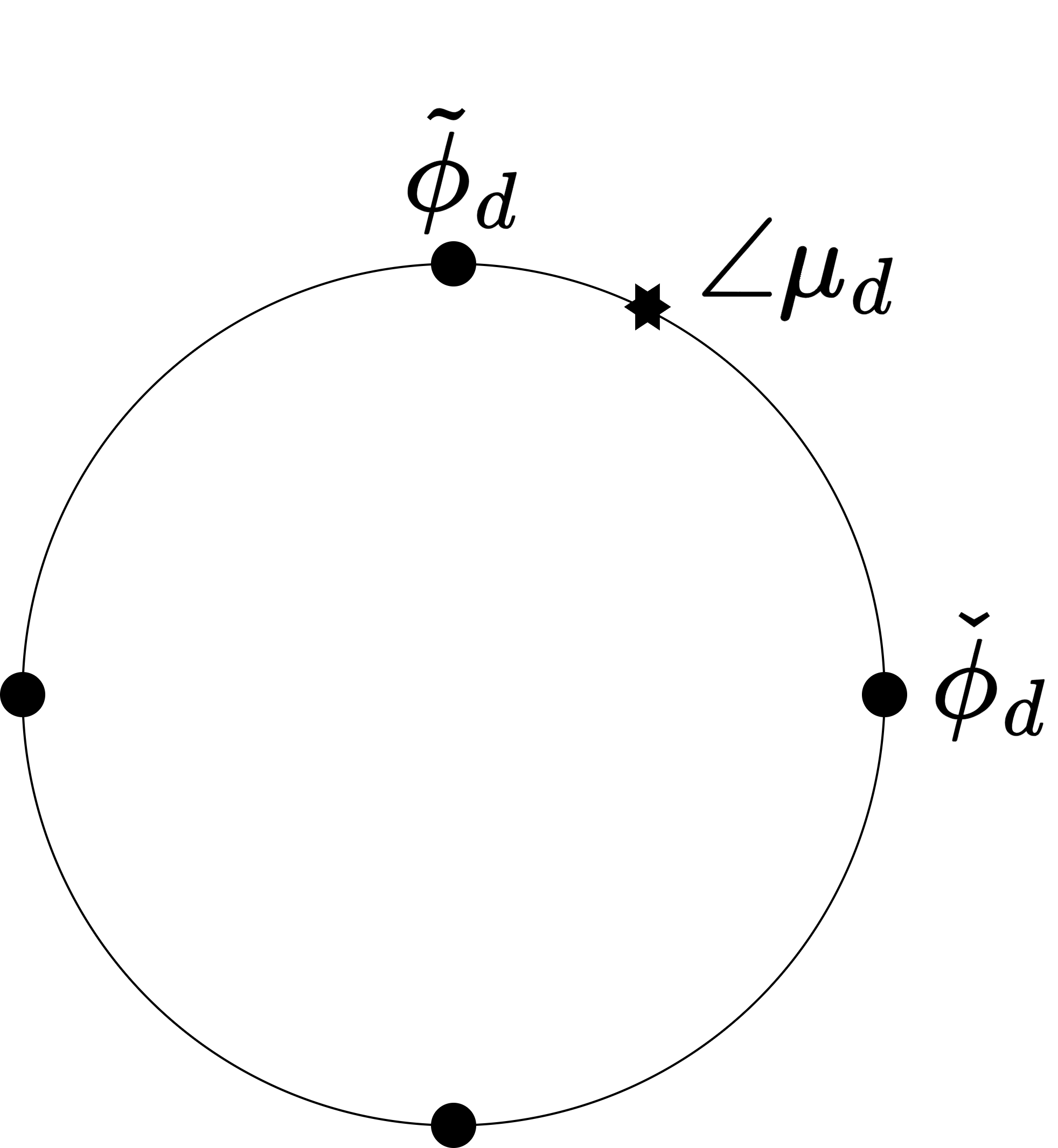}
    \caption{{Closest PSK symbol $\tilde{\phi}_d$ and second closest PSK symbol $\check{\phi}_d$ to $\angle \mu_d$ of the $d$-th undetected symbol.}} 
    \label{fig:choosing_the_closest_phase}
\end{figure}

{
To explain the detection of a $d$-th undetected PSK symbol given all the detected PSK symbols in $\bar{\mathcal{D}}$, let $\bar{\mathbf{v}}_d = [{v}_{k_0},\dots,{v}_{k_{n-1}},{v}_d]^T$, $\tilde{\mathbf{u}}_d = [\tilde{u}_{k_0},\dots,\tilde{u}_{k_{n-1}},\tilde{u}_d]^T$ and $\bar{\mathbf{Y}}_d = [\mathbf{y}_{k_0},\dots,\mathbf{y}_{k_{n-1}},\mathbf{y}_d]$ be transmitted signal, amplitude vector of the transmitted signal and the received signal corresponding to all the detected symbols in $\bar{\mathcal{D}}$ and the $d$-th undetetected symbol. Please note that $\mathbf{y}_i \in \mathbb{C}^{M\times 1}$ be the $i$-th column of $\mathbf{Y}$, which corresponds to the received signal of the $i$-th symbol $v_i$. Then, one can write the PDF of $\bar{\mathbf{Y}}_d$ given $\bar{\mathbf{v}}_d$ \cite{hochwald2000unitary}:}
\begin{align}
    & f(\bar{\mathbf{Y}}_d|\bar{\mathbf{v}}_d) = f(\bar{\mathbf{Y}}_d|\tilde{\mathbf{u}}_d,\bar{\phi}_{{k}_0},\dots, \bar{\phi}_{{k}_{n-1}},\phi_d)  \nonumber \\
    & = \frac{\exp\left(-\frac{\text{tr}(\bar{\mathbf{Y}}_d^H\bar{\mathbf{Y}}_d)}{\sigma^2} + \frac{\text{tr}\left( \bar{\mathbf{Y}}_d^H \bar{\mathbf{Y}}_d \ \bar{\mathbf{v}}_d^* \bar{\mathbf{v}}_d^T \right)}{\sigma^2(\sigma^2 + \Vert \bar{\mathbf{v}}_d \Vert^2)} \right)}{\pi^{(n+1)M}\left[ (\sigma^2 + \Vert \bar{\mathbf{v}}_d \Vert^2)\sigma^{2n} \right]^{M}}.
    \label{eqn:density_func2}
\end{align}
Please note that $\bar{\mathbf{Y}}_d$, $\bar{\phi}_{{k}_0},\dots, \bar{\phi}_{{k}_{n-1}}$ are known to the receiver and $\tilde{\mathbf{u}}_d$ can be found by solving the amplitude detection in \eqref{eqn:u_ML_detector}; hence, only $\phi_d$ is unknown to the receiver. Thus, we simplify (\ref{eqn:density_func2}) as a function of $\phi_d$ as follows:
\begin{align}
    f(\bar{\mathbf{Y}}_d|\tilde{\mathbf{u}}_d,\bar{\phi}_{{k}_0},\dots, \bar{\phi}_{{k}_{n-1}},\phi_d) = \bar{c}_{1} \exp\left(\bar{c}_{2} \text{Re}\left\{ \mu_d e^{-j{\phi}_{d}} \right\}\right),
    \label{eqn:simplified_density_func2}
\end{align}
where $\bar{c}_{1}$, $\bar{c}_{2}$, and $\mu_d$ are constant and independent of $\phi_d$. The value of $\mu_d$ is given as:
\begin{align}
    \mu_d = \sum_{{k} \in \bar{\mathcal{D}}} z^{(u)}_{{k},d} e^{j\bar{\phi}_{{k}}},
    \label{eqn:mu_d}
\end{align}
where $z^{(u)}_{i,j}=\mathbf{y}_{i}^H\mathbf{y}_{j} \tilde{u}_i \tilde{u}_j$ is the $(i,j)$-th entry of $\mathbf{Z}^{(u)}=(\mathbf{Y}^H\mathbf{Y})(\tilde{\mathbf{u}}\tilde{\mathbf{u}}^T)$. The proof of (\ref{eqn:simplified_density_func2}) and the formulas of $\bar{c}_{1}$ and $\bar{c}_{2}$ are provided in the Appendix  \ref{subsection:proof_PDF}. A temporary value  of $\phi_d$ is obtained by solving the following ML detection as follows:
\begin{align}
    & \tilde{\phi}_{d} = \argmax_{{\phi}_{d} \in \Omega_{\phi_d}} \ f(\bar{\mathbf{Y}}_d|\tilde{\mathbf{u}}_d,\bar{\phi}_{{k}_0},\dots, \bar{\phi}_{{k}_{n-1}},\phi_d) \nonumber \\
    & = \argmax_{{\phi}_{d} \in \Omega_{\phi_d}} \ \text{Re}\left\{ \mu_d e^{-j{\phi}_{d}} \right\} = \argmax_{{\phi}_{d} \in \Omega_{\phi_d}} \ |\mu_d| \cos(\angle \mu_d - {\phi}_{d}).
    \label{eqn:decision_DFDD}
\end{align}
Please note that $\angle \mu_d$ is considered to be the reference phase, and the solution $\tilde{\phi}_{d}$ of (\ref{eqn:decision_DFDD}) is obtained by choosing the $d$-th PSK symbol that is closest to $\angle \mu_d$, as shown in Fig. \ref{fig:choosing_the_closest_phase}. The ML detection of the phase in \eqref{eqn:decision_DFDD} can be obtained by an exhaustive search because the size of $\Omega_{\phi_d}$ is typically small.

After the temporary decision of $\phi_d$, i.e., $\tilde{\phi}_{d}$, is obtained, the reliability of the $d$-th symbol can be obtained. {In \cite{fischer2014noncoherent}, the sort-DFDD uses the LLR as the reliability criterion; however, it was limited only to multi-symbol  4-PSK modulations with equal amplitudes.
Our proposed PR-sort-DFDD can detect multi-symbols $2^{l_\phi}$-ary PSK modulations with varying amplitudes.
We define and use the posterior probability as the reliability criterion $R_d$ of the $d$-th symbol as:}
\begin{align}
    \label{eqn:posterior_prob1}
    R_d & = P\left(\tilde{\phi}_{d} | \bar{\mathbf{Y}}_d, \tilde{\mathbf{u}}_d, \bar{\phi}_{{k}_0}, \dots, \bar{\phi}_{{k}_{n-1}}\right), \nonumber \\
    & = \frac{f\left( \bar{\mathbf{Y}}_d | \tilde{\mathbf{u}}_d, \bar{\phi}_{{k}_0}, \dots, \bar{\phi}_{{k}_{n-1}}, \tilde{\phi}_{d} \right)} { \sum_{{\phi}_{d} \in \Omega_{\phi_d} } f\left(\bar{\mathbf{Y}}_d | \tilde{\mathbf{u}}_d, \bar{\phi}_{{k}_0}, \dots, \bar{\phi}_{{k}_{n-1}}, {\phi}_{d} \right)}, \nonumber \\
    & = \frac{ \exp \left( \bar{c}_{2} \left| \mu_d \right| \cos(\angle \mu_d - \tilde{\phi}_{d})\right)}{ \sum_{{\phi}_{d} \in \Omega_{\phi_d} } \exp \left(\bar{c}_{2} \left| \mu_d \right| \cos(\angle \mu_d - {\phi}_{d}) \right) }.
\end{align}
{To simplify the calculation of $R_d$, we consider the two dominant terms in the denominator of (\ref{eqn:posterior_prob1}),
which correspond to the closest and second closest PSK symbols to $\angle \mu_d$, i.e., $\tilde{\phi}_d$ and $\check{\phi}_d$, respectively, as shown in Fig. \ref{fig:choosing_the_closest_phase}. 
That said, the posterior probability in (\ref{eqn:posterior_prob1}) is approximated as follows:}
\begin{align} \label{eqn:posterior_prob1_approx}
    R_d & \approx \frac{e^{  \bar{c}_{2,d} \left|\mu_d\right| \cos(\angle \mu_d - \tilde{\phi}_{d}) } }{ e^{ \bar{c}_{2,d} \left| \mu_d \right| \cos(\angle \mu_d - \tilde{\phi}_{d}) } + e^{ \bar{c}_{2,d} \left| \mu_d \right| \cos(\angle \mu_d - \check{\phi}_{d}) } }, \nonumber \\
    & = \frac{1}{1 + e^{ - \bar{c}_{2,d} \left| \mu_d \right| \left( \cos(\angle \mu_d - \tilde{\phi}_{d}) -  \cos(\angle \mu_d - \check{\phi}_{d}) \right) } }.
\end{align}
Since $1/(1+e^{-x})$ is an increasing function of $x$, we can further simplify $R_d$ without affecting the solution of selecting the $d$-th PSK symbol with the highest $R_d$ as follows:
\begin{align}
    R_d & = \left| \mu_d \right| \left( \cos(\angle \mu_d - \tilde{\phi}_{d}) -  \cos(\angle \mu_d - \check{\phi}_{d}) \right) \nonumber \\
    & = \text{Re}\left\{ \mu_d e^{-j\tilde{\phi}_{d}} \right\} - \text{Re}\left\{ \mu_d e^{-j\check{\phi}_{d}} \right\}.
    \label{eqn:MAP-reliability}
\end{align}
{The PR-sort-DFDD selects the most reliable symbol, $k_n = \argmax_{d \in \mathcal{D}} R_d$, and  we detect the $k_n$-th symbol using (\ref{eqn:decision_DFDD}) and obtain the final decision $\bar{\phi}_{k_n}$ of the $k_n$-symbol. The PR-sort-DFDD then updates the detected set $\bar{\mathcal{D}}$ and the undetected set ${\mathcal{D}}$ by adding $k_n$ to $\bar{\mathcal{D}}$ and removing $k_n$ from ${\mathcal{D}}$. The process continues until all the symbols are detected. The proposed PR-sort-DFDD algorithm is summarized in Algorithm \ref{alg:PR-sort-DFDD}.}
\begin{algorithm}
    \caption{PR-sort-DFDD algorithm}
    {\textbf{Input:} $\mathbf{Z}^{(u)}$, $\bar{\mathcal{D}}=\{0\}$, $\bar{\phi}_0 = 0$, ${\mathcal{D}}=\{1,\dots,K-1\}$. \\
    \textbf{Step 1:} Calculate $\mu_d$ from (\ref{eqn:mu_d}) for each $d \in \mathcal{D}$. \\
    \textbf{Step 2:} Find $\tilde{\phi}_d$ and $\check{\phi}_d$ from (\ref{eqn:decision_DFDD}) for each $d \in \mathcal{D}$. \\
    \textbf{Step 3:} Calculate reliability $R_d$ as in (\ref{eqn:MAP-reliability}) for each $d \in \mathcal{D}$. \\
    \textbf{Step 4:} Select the undetected symbol with the highest $R_d$, i.e., $k_n$ and obtain $\bar{\phi}_{k_n}$ as the final solution of (\ref{eqn:decision_DFDD}). \\
    \textbf{Step 5:} Remove $k_n$ from $\mathcal{D}$ and add $k_n$ to $\bar{\mathcal{D}}$. \\
    \textbf{Step 6:} Repeat \textbf{Step 1} to \textbf{Step 5} until $\mathcal{D}$ is empty. \\
    \textbf{Output:} $\bar{\phi}_0, \dots, \bar{\phi}_{K-1}$.}
    \label{alg:PR-sort-DFDD}
\end{algorithm}

\vspace{-10pt}
\subsection{Detection of Phase Vector with Unequal Modulation Orders}
\label{subsection:improved_sorted_DFDD}

\begin{figure}
    \begin{subfigure}{0.5\textwidth}
        \centering
        \includegraphics[width=5cm,height=1.5cm]{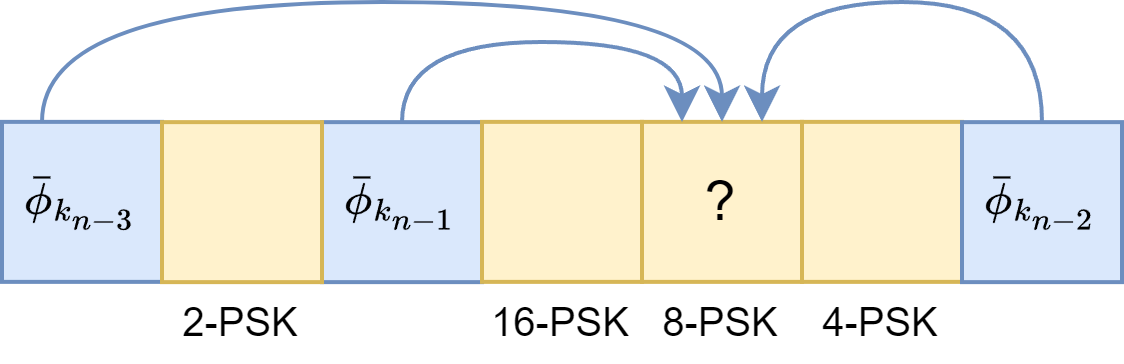}
        \caption{Proposed PR-sort-DFDD}
        \vspace{15pt}
        \label{fig:sortDFDD_mechanism}
    \end{subfigure}
    
    \begin{subfigure}{0.5\textwidth}
        \centering
        \includegraphics[width=5cm,height=1.5cm]{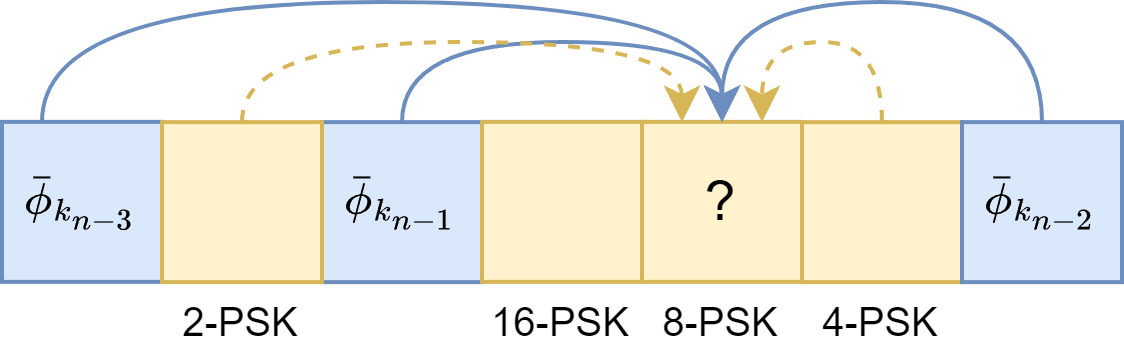}
        \caption{Proposed improved PR-sort-DFDD}
        \label{fig:improved_sortDFDD_mechanism}
    \end{subfigure}
    \caption{Comparison between two proposed algorithms. Blue and yellow boxes indicate detected and undetected symbols, respectively. Blue solid arrows and yellow dashed arrows indicate information from the detected symbol and the undetected symbol, respectively.}
    \label{fig:comparison_sortDFDD}
\end{figure}


{In this subsection, we propose the improved-PR-sort-DFDD algorithm, which improves the performance of the PR-sort-DFDD algorithm in the case of unequal modulation orders of the phase vector $\mathbf{p}$. The improvement in the performance is achieved by exploiting the information from undetected symbols with lower modulation order to enhance the detection of an undetected symbol with higher modulation order. For example, Fig. \ref{fig:improved_sortDFDD_mechanism} shows that our proposed improved-PR-sort-DFDD uses information from undetected 2-PSK, 4-PSK symbols to detect the 8-PSK symbol. Meanwhile, the proposed PR-sort-DFDD (shown in Fig. \ref{fig:sortDFDD_mechanism}) as well as other DFDD algorithms only uses information from detected symbols to detect undetected symbols.}

\begin{figure}[t]
    \begin{subfigure}{0.5\textwidth}
        \centering
        \includegraphics[width=5cm]{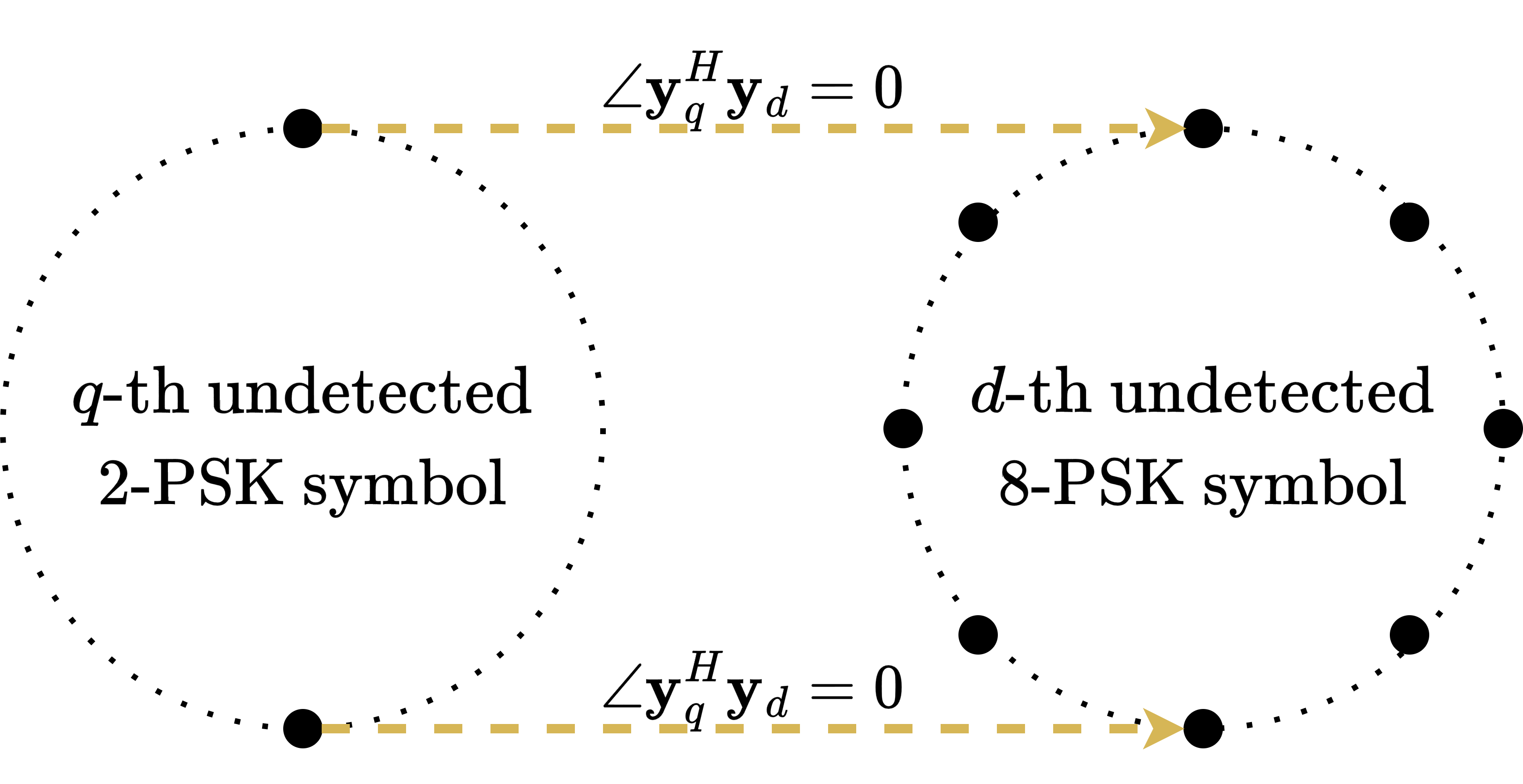}
        \caption{Unequal modulation orders: The phase difference suggests that the $d$-th symbol is more likely to be $\pi/2$ or $-\pi/2$ than other values.} \vspace{20pt}
        \label{fig:unequal_modulation_order_undetected_symbol_to_undetected_symbol}
    \end{subfigure}
    \vfill
    \begin{subfigure}{0.5\textwidth}
        \centering
        \includegraphics[width=5cm]{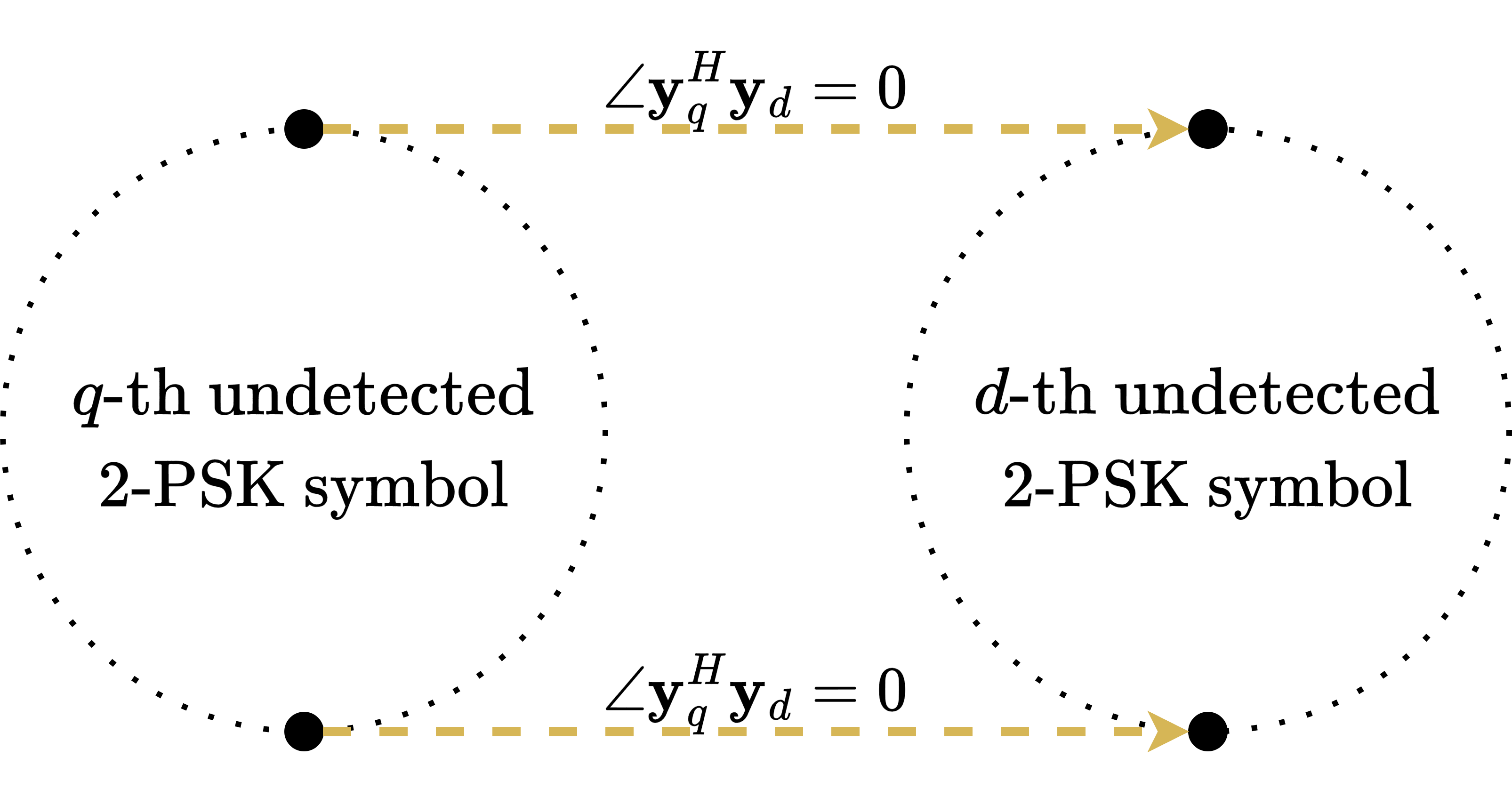}
        \caption{Equal modulation orders: The phase difference suggests that the $d$-th symbol is more likely to be $\pi/2$ or $-\pi/2$. This information is not meaningful because $d$-th symbol has only two possible values.}
        \label{fig:equal_modulation_order_undetected_symbol_to_undetected_symbol}
    \end{subfigure}
    \caption{Illustration of how the phase difference $\angle \mathbf{y}_q^H\mathbf{y}_d$ between two undetected PSK symbols can affect the detection of the $d$-th undetected PSK symbol. We assume that $\angle \mathbf{y}_q^H\mathbf{y}_d=0$.}
    \label{fig:illustration_symbol_affect_symbol}
\end{figure}

{To explain the improved-PR-sort-DFDD, we first give an intuition on how the phase difference between undetected PSK symbols with unequal modulation orders can provide useful information. Fig. \ref{fig:unequal_modulation_order_undetected_symbol_to_undetected_symbol} shows the $q$-th undetected PSK symbol and $d$-th undetected PSK symbol with modulation orders of 2 and 8, respectively. Their phase difference can be estimated by calculating the phase between their corresponding received signal, i.e., $\angle \mathbf{y}_q^H \mathbf{y}_d$. For the sake of simplicity of explaining the concept, let us assume that $\angle \mathbf{y}_q^H \mathbf{y}_d = 0$. Given the phase difference $\angle \mathbf{y}_q^H \mathbf{y}_d = 0$, the $d$-th symbol is most likely to be $\pi/2$ if the $q$-th symbol is $\pi/2$; and the $d$-th symbol is most likely to be $-\pi/2$ if the $q$-th symbol is $-\pi/2$. Because we do not know the correct decision of the $q$-th PSK symbol, we do not know which phase value of the $d$-th PSK symbol is most likely to be. However, we can expect that the $d$-th PSK symbol phase is more likely to be $\pi/2$ or $-\pi/2$ when compared to other possible phase values. This information can be useful in enhancing the detection of the $d$-th symbol. Please note that if the undetected $q$-th and $d$-th PSK symbols have equal modulation orders as shown in Fig. \ref{fig:equal_modulation_order_undetected_symbol_to_undetected_symbol}, then their phase difference only suggests the $d$-th symbol is more likely to be $\pi/2$ or $-\pi/2$. However, the $d$-th symbol is already likely to be $\pi/2$ or $-\pi/2$ without knowing the phase difference because there are only two possible choices of the $d$-th PSK symbol. As a result, the phase difference between two undetected symbols with equal modulation orders does not bring useful information.}

In the following, we investigate the detection of the $d$-th undetected PSK symbol given the observations of the $q$-th and $d$-th undetected PSK symbols. Let ${\mathbf{v}}_{d,q} = [v_d,v_q]$, $\tilde{\mathbf{u}}_{d,q} = [\tilde{u}_d,\tilde{u}_q]$ and ${\mathbf{Y}}_{d,q} = [\mathbf{y}_d,\mathbf{y}_q$] be transmitted signal, amplitude vector of transmitted signal and received signals corresponding to the $d$-th and $q$-th symbol, respectively. The PDF of ${\mathbf{Y}}_{d,q}$ given $\mathbf{v}_{d,q}$,  is given as follows \cite{hochwald2000unitary}:
\begin{align}
    & f(\mathbf{Y}_{d,q}|\mathbf{v}_{d,q}) = f({\mathbf{Y}}_{d,q} | \tilde{\mathbf{u}}_{d,q}, {\phi}_{d}, {\phi}_{{q}} ) .
    \label{eqn:PDF_undetected_symbols}
\end{align}
Since ${\mathbf{Y}}_{d,q}$ is known to the receiver and $\tilde{\mathbf{u}}_{d,q}$ is obtained by the amplitude detection in \eqref{eqn:u_ML_detector}, while $\phi_d$ and $\phi_q$ are unknown, we simplify (\ref{eqn:PDF_undetected_symbols}) as a function of $\phi_d$ and $\phi_q$ as follows:
\begin{align}
    f({\mathbf{Y}}_{d,q} | \tilde{\mathbf{u}}_{d,q}, {\phi}_{d}, {\phi}_{{q}} ) = c_1 \exp \left( c_2 \text{Re} \left\{ z^{(u)}_{d,{q}} e^{j({\phi}_{d}-{\phi}_{{q}})} \right\} \right),
    \label{eqn:PDF_undetected_symbols_simplified}
\end{align}
where $c_1$, $c_2$ and $z^{(u)}_{d,{q}}$ are constant and independent of $\phi_d$ and $\phi_q$. The proof of (\ref{eqn:PDF_undetected_symbols_simplified}) is similar to the proof of (\ref{eqn:simplified_density_func2}). Then, the log of the posterior probability of the $d$-th symbol given all observations regarding the $d$-th symbol and $q$-th symbol, i.e., $\mathbf{Y}_{d,q}$, $\tilde{u}_d$ and $\tilde{u}_q$, is given by:
\begin{align}
    \label{eqn:log_likelihood_given_other_symbols}
    & \log \left( P( {\phi}_{d} | {\mathbf{Y}}_{d,q}, \tilde{{u}}_{d}, \tilde{{u}}_{q} ) \right) \nonumber \\
    & \propto \log \left( \sum_{{\phi}_{{q}} \in \Omega_{\phi_{{q}}}} \ f({\mathbf{Y}}_{d,q} | \tilde{u}_d,\tilde{u}_q, {\phi}_{d}, {\phi}_{{q}} ) \right), \nonumber \\
    & \approx \log \left( \max_{{\phi}_{{q}} \in \Omega_{\phi_{{q}}}} \ c_1 \exp \left( c_2 \text{Re} \left\{ z^{(u)}_{d,{q}} e^{j({\phi}_{d}-{\phi}_{{q}})} \right\} \right) \right), \nonumber \\
    & \propto  \eta_{d,q}(\phi_d),
\end{align}
The approximation in \eqref{eqn:log_likelihood_given_other_symbols} is achieved by the {Max-Log approximation}. The term $\eta_{d,q}(\phi_d)$ is defined as:
\begin{align}
    \label{eqn:max_correlation}
    \eta_{d,q}(\phi_d) = \max_{{\phi}_{{q}} \in \Omega_{\phi_{{q}}}} \text{Re}\left\{ z^{(u)}_{{d},q} e^{j({\phi}_{{d}}-{\phi}_{q})} \right\}.
\end{align}
Since $\eta_{d,q}(\phi_d)$ in \eqref{eqn:max_correlation} represents the log-likelihood of the $d$-th PSK symbol given the observations of both the $d$-th and $q$-th PSK symbols, we include it as the selection criterion of the symbol to be detected in the PR-sort-DFDD; and this new algorithm is the improved-PR-sort-DFDD. The decision rule in (\ref{eqn:decision_DFDD}) is modified as follows:
\begin{align}
    \tilde{\phi}_d = \argmax_{\phi_d \in \Omega_{\phi_d}}  \underbrace{\text{Re}\left\{ \mu_d e^{-j{\phi}_{d}} \right\}}_{\substack{\text{log-likelihood given} \\ \text{all detected symbols}} } + \sum_{{q} \in \mathcal{D}/\{ d \}, l_{\phi_{{q}}} < l_{\phi_{d}}} \underbrace{\eta_{d,q}(\phi_d)}_{\substack{\text{log-likelihood given} \\ \text{$q$-th undetected symbol}}}.
    \label{eqn:modified_DFDD_func}
\end{align}
We follow a similar approach as PR-sort-DFDD in (\ref{eqn:MAP-reliability}) by taking $\check{\phi}_{k_n}$ as a solution with the second-highest value of (\ref{eqn:modified_DFDD_func}) and calculate the reliability criterion based on the difference of likelihood between $\tilde{\phi}_d$ and $\check{\phi}_{k_n}$, which is given by:
\begin{align}
    R_d = & \text{Re}\left\{ \mu_d e^{-j\tilde{\phi}_{d}}\right\} - \text{Re}\left\{ \mu_d e^{-j\check{\phi}_{d}}\right\} \nonumber \\ 
    & + \sum_{{q} \in \mathcal{D}/\{ d \}, l_{\phi_{{q}}} < l_{\phi_{d}}} \eta_{d,q}(\tilde{\phi}_d) - \eta_{d,q}(\check{\phi}_d) .
    \label{eqn:kn_decision_modified_DFDD}
\end{align}
{The improved-PR-sort-DFDD algorithm is similar to the PR-sort-DFDD in Alg. \ref{alg:PR-sort-DFDD}, except that $\tilde{\phi}_d$ and $\check{\phi}_d$ are calculated from (\ref{eqn:modified_DFDD_func}) and $R_d$ is calculated from (\ref{eqn:kn_decision_modified_DFDD}) in the improved-PR-sort-DFDD algorithm.} Please note that if we use the improved-PR-sort-DFDD for equal modulation orders, the decision rule (\ref{eqn:modified_DFDD_func}) and reliability criterion (\ref{eqn:kn_decision_modified_DFDD}) of the improved-PR-sort-DFDD algorithm is reduced to the decision rule (\ref{eqn:decision_DFDD}) and reliability criterion (\ref{eqn:MAP-reliability}) of the PR-sort-DFDD algorithm.


\subsection{Detection Complexity Analysis}
\label{subsection:detection_complexity}
{Overall, the detection process includes calculating $\mathbf{Y}^H\mathbf{Y}$, detecting the amplitude vector $\mathbf{u}$, and the phase vector $\mathbf{p}$.}

{The complexity of the amplitude vector detection in \eqref{eqn:u_ML_detector}, is $\mathcal{O}(2^{l_u}K^2)$. This is as calculating the argument in \eqref{eqn:u_ML_detector} requires $\mathcal{O}(K^2)$ and solving the optimization problem in \eqref{eqn:u_ML_detector} for all possible $2^{l_u}$ amplitude vectors requires  an exhaustive search of complexity order $\mathcal{O}(2^{l_u})$. }

{Regarding the complexity of the phase vector detection, the PR-sort-DFDD and the improved-PR-sort-DFDD include two nested loops with a maximum of $K^2$ iterations. In case of PR-sort-DFDD, each iteration includes the computation of (\ref{eqn:decision_DFDD}) with complexity of $\mathcal{O}(K + 2^{l_{\phi_d}})$.
Thus, the worst-case complexity of the PR-sort-DFDD is $\mathcal{O}(K^3+ 2^{l_{\phi_{\max}}} K^2)$. Compared to PR-sort-DFDD, the improved-PR-sort-DFDD have higher complexity due to the calculation of $\eta_{d,q}(\phi_d)$ in (\ref{eqn:modified_DFDD_func}). Please note that $\eta_{d,q}$ can be calculated and stored before starting the improved-PR-sort-DFDD, and the worst-case complexity of pre-calculating every possible values of $\eta_{d,q}(\phi_d)$ is $2^{l_{\phi_{\max}}}K^2$. Hence, the complexity of improved-PR-sort-DFDD is still $\mathcal{O}(K^3+ 2^{l_{\phi_{\max}}} K^2)$.}

{Given the complexity of calculating $\mathbf{Y}^H\mathbf{Y}$ is $\mathcal{O}(MK^2)$, the overall computational complexity order of the proposed IUAP  is $\mathcal{O}\left(MK^2 + 2^{l_u} K^2 + K^3 + 2^{l_{\phi,\text{max}}} K^2 \right)$ excluding the number of iterations required for the IUAP algorithm to converge. As can be seen, the complexity of the proposed IUAP algorithm is polynomial in $K$ and exponential in $l_u$.
}

\begin{table}[t]
    \centering
    \caption{{Comparison of detection complexity for $M \geq K$.} }\vspace{-10pt}
    \begin{tabular}{|c|c|c|c|}
        \hline
        Detectors & Detection complexity\\
        \hline
         ML detector & $\mathcal{O} \left( MK^2 + 2^{l_v}K^2  \right)$ \\
        \hline
        Exponential/cube-split & $\mathcal{O}(MK^2)$  \\
        \hline
        Proposed detectors & $\mathcal{O}(MK^2 + 2^{l_u} K^2 + K^3 + 2^{l_{\phi,\text{max}}} K^2 )$ \\
        \hline
    \end{tabular}
    \label{tab:complexity_comparison}
\end{table}

{In Table \ref{tab:complexity_comparison}, we compare the complexity of our proposed detectors for proposed constellations with the detection complexity of other unitary constellations in the literature, assuming that $M$ is large and $K$ is small; hence, $M\geq K$. Since $2^{l_u} + 2^{l_{\phi_{\max}}} \ll 2^{l_v}$ as shown in Section \ref{subsection:bit_allocation}, our proposed detectors have significantly lower complexity than the ML detector. However, our proposed detectors for our proposed constellations have higher detection complexity than the detectors of exponential or cube-split constellations.} 

\section{Simulation Result}
\label{section:simulation_result}

In this section, we evaluate the MCD and the error performance achieved by our proposed unitary constellation with proposed detectors and compare them with other schemes.
Unless otherwise mentioned, the number of antennas is set to $M = 32$. Please note that we evaluate the performance for $K \geq 3$ since our proposed unitary constellation has the same structure as the one in \cite{li2019design} in the case of $K=2$.

\begin{figure}[t]
    \begin{subfigure}{0.45\textwidth}
        \centering
        \includegraphics[height=2.5cm,width=8cm]{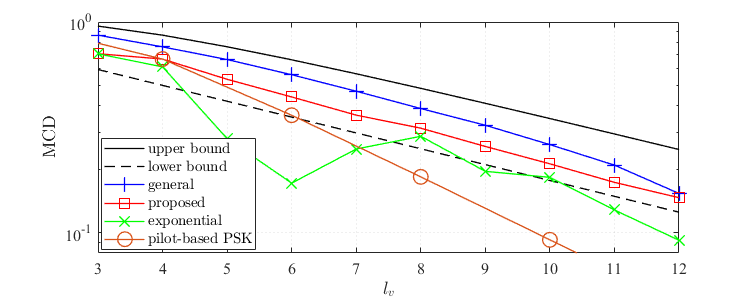}
        \caption{$K=3$} \vspace{10pt}
        \label{fig:chordal_distance_vs_lv_Ts3}
    \end{subfigure}
    \hfill
    \begin{subfigure}{0.45\textwidth}
        \centering
        \includegraphics[height=2.5cm,width=8cm]{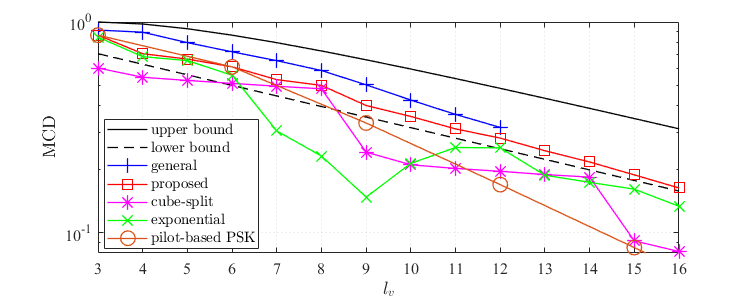}
        \caption{$K=4$} 
        \label{fig:chordal_distance_vs_lv_Ts4}
    \end{subfigure}
    \caption{MCD comparison for different values of $l_v$ and $K$.}
    \label{fig:chordal_distance_vs_lv}
\end{figure}

Fig. \ref{fig:chordal_distance_vs_lv} depicts the MCD of the proposed unitary constellation with the following unitary constellations: general constellation \cite{hochwald2000unitary}, exponential constellation \cite{kammoun2007non}, cube-split constellation \cite{ngo2019cube}, and the pilot-based PSK constellations which include one pilot and $K-1$ PSK symbols with equal amplitude and equal modulation orders. 
As can be seen from Fig. \ref{fig:chordal_distance_vs_lv}, the MCD of the proposed unitary constellation is higher, hence better, than its counterparts of the exponential, the cube-split, and the pilot-based PSK constellations, and it is close to the MCD of the general unitary constellation for $K = $ 3, 4.

\begin{figure}[t]
    \centering
    \includegraphics[height=5.2cm,width=8cm]{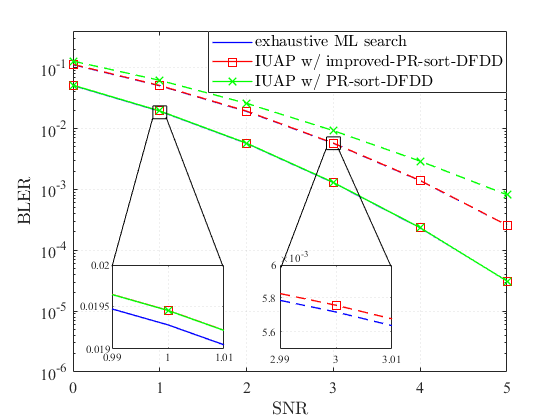}
    \caption{Performance comparisons with ML detector. Solid lines and dashed lines correspond to equal modulation orders ($\{ l_{\phi_1},l_{\phi_2},l_{\phi_3}\}=\{2,2,2\}$) and unequal modulation orders ($\{ l_{\phi_1},l_{\phi_2},l_{\phi_3}\}=\{2,2,3\}$), respectively.}
    \label{fig:BLER_different_detection_schemes}
\end{figure}


We compare the BLER of our proposed IUAP with PR-sort-DFDD and IUAP with improved-PR-sort-DFDD, with the optimal ML detector in (\ref{eqn:GLRT_detector}), obtained through an exhaustive search. Please note that since the amplitude set $\Omega_u$ does not have a specific structure, the amplitude detection of the IUAP is carried out by an exhaustive search. As can be seen from Fig. \ref{fig:BLER_different_detection_schemes}, both proposed detectors achieve the performance of the ML detector for the case of equal modulation orders with negligible 0.01 dB SNR penalty; while only the performance of the IUAP with the PR-sort-DFDD deteriorates for the case of unequal modulation orders.


\begin{figure}[t]
    \begin{subfigure}{0.45\textwidth}
        \centering
        \includegraphics[height=5.2cm,width=8cm]{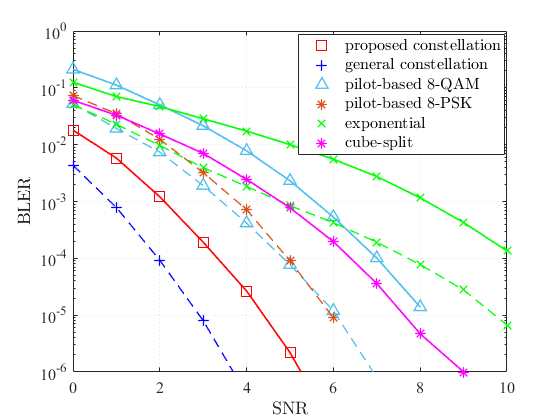}
        \caption{$K=4$, $l_v=9$}
        \label{fig:BLER_vs_SNR_Ts4_9bits_updated}
    \end{subfigure}
    \caption{BLER comparisons of different schemes. {Dashed lines and solid lines corresponds to high-complexity ML detectors and low-complexity detectors, respectively.}}
    \label{fig:BLER_vs_SNR_32antennas_B0=3}
\end{figure}

In the following, we compare the error performance of our proposed unitary constellations with general constellation \cite{hochwald2000unitary}, exponential constellation \cite{kammoun2007non}, cube-split constellation \cite{ngo2019cube}, and the pilot-based schemes for different values of $K$ and $l_v$. The pilot-based schemes include one pilot and $K-1$ QAM or PSK symbols. The optimal ML detector in (\ref{eqn:GLRT_detector}) is required for the general constellations, while low-complexity detectors can be used for proposed unitary constellations (IUAP with improved-PR-sort-DFDD), cube-split constellations (greedy detector), exponential constellations (simplified detector), and the pilot-based QAM scheme (MMSE channel estimation and MMSE linear detector). Moreover, to show the best potential error performance of all the schemes, we simulate the ML detector for all the schemes except for the proposed unitary constellation and the cube-split constellations, as their low-complexity detectors already have near-optimal performance of the ML detector. Please note that the ML detector's performance does not necessarily reflect the actual performance in practical applications due to its prohibitive complexity.

\begin{figure}[t]
    \begin{subfigure}{0.5\textwidth}
        \centering
        \includegraphics[height=5.2cm,width=8cm]{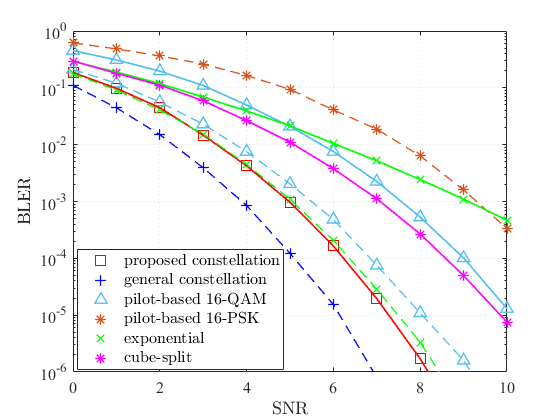}
        \caption{BLER}\vspace{10pt}
        \label{fig:BLER_vs_SNR_Ts4_12bits_updated}
    \end{subfigure}
    \begin{subfigure}{0.5\textwidth}
        \centering
        \includegraphics[height=5.2cm,width=8cm]{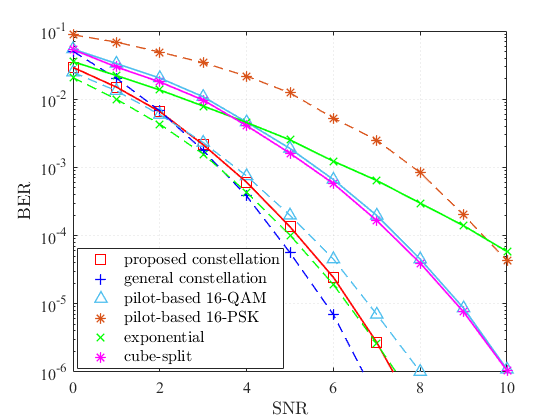}
        \caption{BER}
        \label{fig:BER_vs_SNR_Ts4_12bits_updated}
    \end{subfigure}
    \caption{BLER and BER comparison of different schemes at $K=4$, $l_v=12$. {Dashed lines and solid lines correspond to high-complexity ML detectors and low-complexity detectors, respectively.}}
    \label{fig:Performance_32antennas_Ts=4_12bit}
\end{figure}

Fig. \ref{fig:BLER_vs_SNR_32antennas_B0=3} and Fig. \ref{fig:Performance_32antennas_Ts=4_12bit} show that our proposed constellation outperforms all schemes with low-complexity detectors by a considerable margin. In specific, our proposed constellations outperform the pilot-based QAM and the cube-split constellations by at least 3 dB and outperforms the exponential constellation by almost 6 dB. Furthermore, our proposed constellation outperforms the pilot-based QAM, pilot-based PSK, and exponential constellations with their ML detector. As expected, the BLER of the general unitary constellation with the ML detector is better than its counterpart of our proposed constellation at different values of $K$ and $l_v$. Please note that for the BER performance reported in Fig. \ref{fig:BER_vs_SNR_Ts4_12bits_updated}, the performance gap between our proposed constellation and the general constellation is slightly less than the BLER performance gap, because the general constellation does not have an efficient mapping criterion due to their random structure. In contrast, our proposed constellation can use Gray mapping to the PSK constellations, which improves its BER.

\begin{figure}[t]
    \centering
    \includegraphics[height=5.2cm,width=8cm]{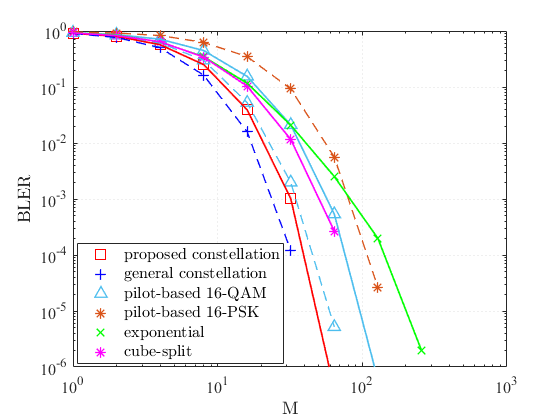}
    \caption{BLER as a function of $M$ at $K=4$, $l_v=12$ and SNR = 5 dB. {Dashed lines and solid lines correspond to high-complexity ML detectors and low-complexity detectors, respectively.}}
    \label{fig:BLER_vs_antennas}
\end{figure}

{
In Fig. \ref{fig:BLER_vs_antennas}, we plot the BLER performance of the proposed unitary constellations and other competing unitary constellations as a function of $M$.
As can be seen, the performance of our proposed unitary constellation improves when the value of $M$ increases. This aligns with the fact that the PEP of unitary constellations decreases as $M$ increases in (\ref{eqn:Chernoff_bound}). Furthermore, Fig. \ref{fig:BLER_vs_antennas} also shows that for a sufficiently large number of antennas (typically $M \geq 8$) our proposed constellations also outperform other low-complexity constellations (exponential and cube-split) and pilot-based QAM with both coherent detectors and non-coherent ML detector.}

\section{Conclusion}
\label{section:conclusion}
In this paper, we proposed a novel design of unitary constellations, which is the Cartesian product of amplitude and phase vectors. The phase vector is confined to a generalized PSK structure where the PSK constellations can have unequal modulation orders and varying amplitude. By exploiting the Cartesian structure, we proposed a low-complexity algorithm that performs amplitude detection and phase detection iteratively. For the phase detection, we adopted the posterior probability as the reliability of the sort-DFDD algorithm, and proposed the PR-sort-DFDD algorithm, which approaches the optimal performance of ML detector in the case of PSK structure with equal modulation orders. For PSK structures with unequal modulation orders, we used feedback from undetected symbols with lower modulation order to enhance the detection of unknown symbols with higher modulation order and proposed the improved-PR-sort-DFDD algorithm. Both of the proposed phase detectors approach the ML detection performance, in their designated cases, with polynomial time complexity. Simulation results showed that the proposed constellations achieve better MCD when compared to other competing low-complexity unitary constellations, such as the exponential and cube-split constellations. For error performance, our proposed constellation with the proposed detectors outperforms conventional pilot-based QAM with coherent detection by 3 dB. Our proposed scheme also outperforms other low-complexity unitary constellations by up to 6 dB.

\vspace{-5pt}
\begin{appendix}

\subsection{Proof of the expression of $D_p(\mathbf{u})$ in (\ref{eqn:closed_form_phase_distance})}
\label{appendix_A1}
First, let $\Delta \mathbf{p} = [e^{j\Delta\phi_0},\dots,e^{j\Delta\phi_{K-1}}]$ with $\Delta\phi_i=\phi_{i,a}-\phi_{i,b}$ being the phase difference between $\mathbf{p}_a$ and $\mathbf{p}_b$. Since the difference of two PSK values $\phi_{i,a}$ and $\phi_{i,b}$ is also a PSK value, hence $\Delta\mathbf{p} \in \Omega_p(\mathbf{l}_{\phi})$. Then, we rewrite $D_p(\mathbf{u})$ as follows:
{\small{
\begin{align}
    D_p(\mathbf{u}) & = \min_{ \substack{\mathbf{p}_a \neq \mathbf{p}_b \\ \mathbf{p}_a,\mathbf{p}_b \in \Omega_p{(\mathbf{l}_{\phi})} } } \sqrt{ 1 - \left| \sum_{i=0}^{K-1} u_{i}^2 e^{j(\phi_{i,b} - \phi_{i,a})} \right|^2 } \nonumber \\
    & = \min_{ \substack{ \Delta \mathbf{p} \in \Omega_p{(\mathbf{l}_{\phi})} \backslash \mathbf{1}_{K} } } \sqrt{ 1 - \left| \sum_{i=0}^{K-1} u_{i}^2 e^{j\Delta \phi_{i}} \right|^2 } .
    \label{eqn:distance_by_phase_difference}
\end{align}
}}
Let $\Delta \hat{\mathbf{p}}$ be the optimal argument of (\ref{eqn:distance_by_phase_difference}). 
If we can find $\Delta \hat{\mathbf{p}}$, we can easily obtain $D_p(\mathbf{u})$ by substituting $\Delta \hat{\mathbf{p}}$ into the objective function (\ref{eqn:distance_by_phase_difference}). One approach to find $\Delta \hat{\mathbf{p}}$ is to solve (\ref{eqn:distance_by_phase_difference}) using exhaustive search over the set $\Omega_p{(\mathbf{l}_{\phi})} \backslash \mathbf{1}_{K}$. However, this approach is not practical because the size of the set $\Omega_p{(\mathbf{l}_{\phi})} \backslash \mathbf{1}_{K}$ grows exponentially with $l_p$. To overcome this problem, we aim to find a subset of $\Omega_p{(\mathbf{l}_{\phi})} \backslash \mathbf{1}_{K}$ (let us say $\Omega^{\ddag}_p \in \Omega_p{(\mathbf{l}_{\phi})} \backslash \mathbf{1}_{K}$) that guarantees to contain the optimal solution $\Delta \hat{\mathbf{p}}$. Thus, $D_p(\mathbf{u})$ can be rewritten as follows:
{\small{
\begin{align}
    D_p(\mathbf{u}) = \min_{ \substack{ \Delta \mathbf{p} \in \Omega^{\ddag}_p } } \sqrt{ 1 - \left| \sum_{i=0}^{K-1} u_{i}^2 e^{j\Delta \phi_{i}} \right|^2 }.
\end{align}
}}
The subset $\Omega^{\ddag}_p$ should have a sufficiently small size so an exhaustive search over $\Omega^{\ddag}_p$ still has low complexity. Such subset $\Omega^{\ddag}_p$ can be given in the Lemma \ref{lemma_0}.

\begin{lemma}
    \label{lemma_0}
    Let $\Omega^{\ddag}_{p,1}$ be a subset consisting of the $Q_1$ elements: $\left\{ \Delta\mathbf{p}^{\ddag}_{k_1}, \dots, \Delta\mathbf{p}^{\ddag}_{k_{Q_1}} \right\}$ corresponding to $\mathcal{K}$, where $\mathcal{K} = \{k_1,k_2,\dots,k_{Q_1}\}$  is a set of the non-zero indices of $\mathbf{l}_{\phi}$. Each element $\Delta\mathbf{p}^{\ddag}_{k} = \left[\Delta {p}^{\ddag}_{0,k}, \dots, \Delta {p}^{\ddag}_{K-1,k}\right]^T$ is given by:
    {\small{
    \begin{align}
            \Delta {p}^{\ddag}_{i,k} = \begin{cases}
                e^{j2\pi/2^{l_{\phi_{k}}}}, \ & \text{if} \ i=k \\
                e^{j0}, \ & \text{if} \ i \neq k
            \end{cases}
            \ , \ i=0,\dots,K-1.
            \label{eqn:sub_phaseSet1} 
    \end{align}
}}
    Let $\Omega^{\ddag}_{p,2}$  be a subset consisting of $Q_2$ elements: $\left\{ \Delta\mathbf{p}^{\ddag}_{l_1}, \dots, \Delta\mathbf{p}^{\ddag}_{l_{Q_2}} \right\}$ corresponding to $\mathcal{L}$, where $\mathcal{L} = \{l_1, l_2, \dots, l_{Q_2}\}$ is a set of the non-zeros indices in ascending order of $\mathbf{l}_{\phi}$ without repetition. Each element $\Delta\mathbf{p}^{\ddag}_{l} = \left[\Delta {p}^{\ddag}_{0,l}, \dots, \Delta {p}^{\ddag}_{K-1,l}\right]^T$ is given by:
    {\small{
    \begin{align}
            \Delta {p}^{\ddag}_{i,l} = \begin{cases}
                e^{j2\pi/2^{l}}, \ & \text{if} \ l_{\phi_i} \geq l \\
                e^{j0}, \ & \text{if} \ l_{\phi_i} < l
            \end{cases}
            \ , \ i=0,\dots,K-1.
            \label{eqn:sub_phaseSet2}
    \end{align}
}}
    Then, a subset $\Omega^{\ddag}_p$ given by $\Omega^{\ddag}_p = \Omega^{\ddag}_{p,1}{(\mathbf{l}_{\phi})} \cup \Omega^{\ddag}_{p,2}{(\mathbf{l}_{\phi})}$ must contain the optimal argument $\Delta \hat{\mathbf{p}}$ of (\ref{eqn:distance_by_phase_difference}).
\end{lemma}

\emph{Proof:} The proof is provided in the Appendix \ref{appendix_AA1}.

Please note that $D^{(k)}_{p,1} (\mathbf{u})$ in (\ref{eqn:first_phase_distance}) is obtained by substituting  $\Delta\mathbf{p}^{\ddag}_{k}$ for $k \in \mathcal{K}$ into (\ref{eqn:distance_by_phase_difference}), and $D^{(l)}_{p,2} (\mathbf{u})$ in (\ref{eqn:second_phase_distance}) is obtained by substituting $\Delta\mathbf{p}^{\ddag}_{l}$ for $l \in \mathcal{L}$ into (\ref{eqn:distance_by_phase_difference}). Thus, we obtain $D_p(\mathbf{u})$ in (\ref{eqn:closed_form_phase_distance}) by substituting all elements of $\Omega^{\ddag}_p$ into (\ref{eqn:distance_by_phase_difference}).

\vspace{-10pt}
\subsection{Proof of the Lemma \ref{lemma_0}}
\label{appendix_AA1}

Let $[\hat{\phi}_0,\dots,\hat{\phi}_{K-1}]$ be the angle of the optimal $\Delta \hat{\mathbf{p}}$, and $\Delta \hat{\phi}_{\text{avg}}$ be the average angle of $\Delta \hat{\mathbf{p}}$, i.e., $\Delta\hat{\phi}_{\rm{avg}} = \angle \left(\sum_{i=0}^{K-1} u_i^2 e^{j\Delta\hat{\phi}_i}\right)$.
Since $\left| \sum_{i=0}^{K-1} u_{i}^2 e^{j\Delta \phi_{i}} \right| = \left| \sum_{i=0}^{K-1} u_{i}^2 e^{-j\Delta \phi_{i}} \right|$, there always exists a pair of $\Delta \hat{\mathbf{p}}$. Thus, for simplicity, we only consider $\Delta \hat{\mathbf{p}}$ with $\Delta\hat{\phi}_{\rm{avg}} \in [0,\pi]$.

We prove Lemma \ref{lemma_0} by contradiction. In specific, if $\Delta \hat{\mathbf{p}}$ lies outside of $\Omega^{\ddag}_p$, we can find an counter example, i.e., $\Delta \check{\mathbf{p}}$, whose objective value in (\ref{eqn:distance_by_phase_difference}) is lower than that of $\hat{u}^2_{\text{avg}}$ of $\Delta \hat{\mathbf{p}}$. Throughout the proof, we denote the complementary subset of $\Omega^{\ddag}_p$ as $\Omega^{\S}_p$, and denote the angle and average angle of $\Delta \check{\mathbf{p}}$ as $[\check{\phi}_0,\dots,\check{\phi}_{K-1}]$ and $\Delta\check{\phi}_{\rm{avg}}$, respectively.

Then, we have the following lemma:
\begin{lemma}
\label{lemma_1}
    If there exists an optimal $\Delta \hat{\mathbf{p}} \in \Omega^{\S}_p$, then
    \begin{align}
        \Delta\hat{\phi}_i = \min_{\Delta\phi_i \in \Omega_{\phi_i}} |\Delta\phi_i - \Delta\hat{\phi}_{\text{avg}}|.
    \end{align}
\end{lemma}
\emph{Proof}: We prove Lemma \ref{lemma_1} by contradiction. Assume that there exists symbol $k$ that satisfies $\Delta\hat{\phi}_k \neq \min_{\Delta\phi_k \in \Omega_{\phi_k}} |\Delta\phi_k - \Delta\hat{\phi}_{\text{avg}}|$. Given $\Delta \hat{\mathbf{p}}$, we choose the counter example $\Delta \check{\mathbf{p}}$ as follows:
\begin{align}
    \Delta\check{\phi}_i =
        \begin{cases}
            \Delta\hat{\phi}_k - \text{sgn}\left(\Delta\hat{\phi}_k - \Delta\hat{\phi}_{\text{avg}}\right) \frac{2\pi}{2^{l_{\phi_k}}}, & \text{if} \ i = k,\\
            \Delta\hat{\phi}_i, & \text{if} \ i \neq k.
        \end{cases}
\end{align}

Let $\{l_0,\dots,l_{Q_2}\}$ contain values of $\mathbf{l}_{\phi}$ in an ascending order. This set is similar to $\mathcal{L}$ defined in (\ref{eqn:closed_form_phase_distance}), except that $\mathcal{L}$ does not contain $l_0=0$. Then, we have the following lemmas:
\begin{lemma}
    There does not exist an optimal $\Delta \hat{\mathbf{p}} \in \Omega^{\S}_p$ with its average phase $\Delta\hat{\phi}_{\text{avg}} \in \left[2\pi / 2^{l_{q+1}}, \ \pi / 2^{l_{q}} \right)$ with $0 \leq q < Q_2$.
    \label{lemma_2}
\end{lemma}
\emph{Proof:}
We prove Lemma~\ref{lemma_2} by contradiction. Assume that $\Delta \hat{\mathbf{p}} \in \Omega^{\S}_p$ with $\Delta\hat{\phi}_{\text{avg}} \in \left[2\pi / 2^{l_{q+1}}, \ \pi / 2^{l_{q}} \right)$. Clearly, we only need to prove the case of $l_{q+1} > l_q + 1$, since $\left[2\pi / 2^{l_{q+1}}, \ \pi / 2^{l_{q}} \right) = \varnothing$ when $l_{q+1} = l_q + 1$. 
Given the range of $\Delta\hat{\phi}_{\text{avg}}$ and using Lemma~\ref{lemma_1}, we obtain $\Delta \hat{\mathbf{p}}$ where $\Delta \hat{\phi}_i = 0$ if $l_{\phi_i} \leq l_{q}$. Then, we choose the counter example $\Delta \check{\mathbf{p}}$ as follows:
\begin{align}
    \Delta\check{\phi}_i = 
    \begin{cases}
        \Delta\hat{\phi}_i = 0 , & \text{if} \ l_{\phi_i} \leq l_{q}, \\
        \Delta\hat{\phi}_i - \left(\frac{2\pi}{2^{l_{q+1}}}\right) , & \text{if} \ l_{\phi_i} \geq l_{q+1}.
    \end{cases}
\end{align}

\begin{lemma}
    There does not exist an optimal $\Delta \hat{\mathbf{p}} \in \Omega^{\S}_p$ with its average phase $\Delta\hat{\phi}_{\text{avg}} \in \left( \pi / 2^{l_{q}},2\pi / 2^{l_{q}} \right)$ with $0 \leq q \leq Q_2$.
    \label{lemma_2b}
\end{lemma}
\emph{Proof:}
We prove Lemma~\ref{lemma_2b} by contradiction. Assume that there exists $\Delta \hat{\mathbf{p}} \in \Omega^{\S}_p$ with $\Delta\hat{\phi}_{\text{avg}} \in \left( \pi / 2^{l_{q}},2\pi / 2^{l_{q}} \right)$. Lemma~\ref{lemma_2b} is true for $q=0$, since $\left( \pi / 2^{l_{q}},2\pi / 2^{l_{q}} \right) = \left( \pi,2\pi \right)$ contradicts $\Delta\hat{\phi}_{\text{avg}} \in \left[ 0, \pi \right]$. Thus, we only need to prove Lemma~\ref{lemma_2b} in case of $q \geq 1$. Given the range of $\Delta\hat{\phi}_{\text{avg}}$ and using Lemma~\ref{lemma_1}, we obtain $\Delta \hat{\mathbf{p}}$ where $\Delta\hat{\phi}_i = 0$ if $l_{\phi_i} \leq l_{q-1}$. Given $\Delta \hat{\mathbf{p}}$, we choose the counter example $\Delta \check{\mathbf{p}}$ as follows:
\begin{align}
    \Delta\check{\phi}_i = 
    \begin{cases}
        \Delta\hat{\phi}_i = 0 , & \text{if} \ l_{\phi_i} \leq l_{q-1} \\
        \frac{2\pi}{2^{l_{q}}} - \Delta\hat{\phi}_i , & \text{if} \ l_{\phi_i} \geq l_{q}.
    \end{cases}
    \label{eqn:counter_example_lemma2_condition2}
\end{align}

\begin{lemma}
    There does not exist an optimal $\Delta \hat{\mathbf{p}} \in \Omega^{\S}_p$ with its average phase $\Delta\hat{\phi}_{\text{avg}} = \pi / 2^{l_q}$ with $0 \leq q \leq Q_2$.
    \label{lemma_2c}
\end{lemma}
\emph{Proof:}
We prove Lemma~\ref{lemma_2c} by contradiction. Assume that $\Delta \hat{\mathbf{p}} \in \Omega^{\S}_p$ with $\Delta\hat{\phi}_{\text{avg}} = {\pi} / {2^{l_q}}$.
We have two possibilities:

1) $q=0$. Thus, $l_q=l_0=0$ and $\Delta\hat{\phi}_{\text{avg}} = 0$. Given $\Delta\hat{\phi}_{\text{avg}} = 0$, we use  Lemma~\ref{lemma_1} to obtain $\Delta \hat{\mathbf{p}}$ with $\Delta\hat{\phi}_i = 0$ if $l_{\phi_i} = l_q = 0$ and $\Delta\hat{\phi}_i = \pi$ if $l_{\phi_i} \neq l_{q} = 0$. We have two cases: 
\begin{itemize}
    \item If $l_1=1$, then the obtained $\Delta \hat{\mathbf{p}}$ is equivalent to:
    \begin{align}
        \Delta\hat{\phi}_i = 
        \begin{cases}
            0, & \text{if} \ l_{\phi_i} < l_1, \\
            \frac{2\pi}{2^{l_1}}, & \text{if} \ l_{\phi_i} \geq l_{1}.
        \end{cases}
    \end{align}
    Thus, $\Delta \hat{\mathbf{p}} \in \Omega^{\ddag}_{p,2}$, contradicting with $\Delta \hat{\mathbf{p}} \in \Omega^{\S}_p$.
    
    \item If $l_1 \geq 2$, we choose the counter example $\Delta \check{\mathbf{p}}$ as follows:
    \begin{align}
        \Delta\check{\phi}_i = 
        \begin{cases}
            0, & \text{if} \ l_{\phi_i} = l_{q} \\
            \frac{2\pi}{2^{l_1}}, & \text{if} \ l_{\phi_i} \neq l_{q}.
        \end{cases}
    \end{align}
\end{itemize}

2) $q \geq 1$. Given $\Delta\hat{\phi}_{\text{avg}} = \pi/2^{l_q}$, we use Lemma \ref{lemma_1} to obtain $\Delta \hat{\mathbf{p}}$ as follows:
\begin{align}
    \Delta\hat{\phi}_i = 
        \begin{cases}
            0 , & \text{if} \ l_{\phi_i} = 0, \\
            0 \text{ or } \frac{2\pi}{ 2^{l_q} }, & \text{if} \ l_{\phi_i} = l_{q}, \\
            \frac{\pi}{2^{l_q}}, & \text{if} \ l_{\phi_i} \neq l_{q}.
        \end{cases}
\end{align}
We choose only one index $k$ where $l_{\phi_k} = l_{q}$, and choose the counter example $\Delta \check{\mathbf{p}}$ as follows:
\begin{align}
    \Delta\check{\phi}_i = 
    \begin{cases}
        \Delta\hat{\phi}_i , & \text{if} \ i \neq k \\
        \frac{2\pi}{2^{l_{\phi_k}}} - \Delta\hat{\phi}_k , & \text{if} \ i = k.
    \end{cases}
    \label{eqn:counter_example_lemma2_condition3}
\end{align}

\begin{lemma}
    There does not exist an optimal $\Delta \hat{\mathbf{p}} \in \Omega^{\S}_p$ with its average phase $\Delta\hat{\phi}_{\text{avg}} \in [0,\pi / 2^{l_{Q_2}})$.
    \label{lemma_3}
\end{lemma}
\emph{Proof:} We prove Lemma~\ref{lemma_3} by contradiction. Assume that there exists $\Delta \hat{\mathbf{p}}\in \Omega^{\S}_p$ with $\Delta\hat{\phi}_{\text{avg}} \in [0,\pi / 2^{l_{Q_2}})$. Given the range of $\Delta\hat{\phi}_{\text{avg}}$, we use Lemma \ref{lemma_1} to obtain $\Delta \hat{\mathbf{p}}$ as follows:
\begin{align}
    \Delta\hat{\phi}_i = \min_{\Delta\phi_i \in \Omega_{\phi_i}} |\Delta\phi_i - \Delta\hat{\phi}_{\text{avg}}| = 0.
\end{align}
Thus, $\Delta \hat{\mathbf{p}} = \mathbf{1}_K$, which is not valid. 

Combining Lemma~\ref{lemma_2}, Lemma~\ref{lemma_2b} and Lemma~\ref{lemma_2c}, we obtain two corollaries:
\begin{corollary}
    There does not exist an optimal $\Delta \hat{\mathbf{p}} \in \Omega^{\S}_p$ with its average phase $\Delta\hat{\phi}_{\text{avg}} \in \left[ 2\pi / 2^{l_{q+1}},2\pi / 2^{l_{q}} \right)$ for $0 \leq q < Q_2$.
    \label{corollary_1}
\end{corollary}

\begin{corollary}
    There does not exist an optimal $\Delta \hat{\mathbf{p}} \in \Omega^{\S}_p$ with its average phase $\Delta\hat{\phi}_{\text{avg}} \in \left[ \pi / 2^{l_{Q_2}},2\pi / 2^{l_{Q_2}} \right)$ for $0 \leq q \leq Q_2$
    \label{corollary_2}
\end{corollary}

Finally, using Corollary~\ref{corollary_1}, Corollary~\ref{corollary_2}, and Lemma~\ref{lemma_3}, we list all possible range of $\Delta\hat{\phi}_{\text{avg}}$, for which an optimal $\Delta \hat{\mathbf{p}} \in \Omega^{\S}_p$ does not exist:
\begin{align}
    \bigcup_{q=0}^{Q_2-1} \left[\frac{2\pi}{2^{l_{q+1}}}, \frac{2\pi}{2^{l_{q}}}\right) \cup \left[\frac{2\pi}{2^{l_{Q_2}}}, \frac{2\pi}{2^{l_{Q_2}}}\right) \cup \left[0, \frac{\pi}{2^{l_{Q_2}}}\right) = \left[0,2\pi\right).
\end{align}
Hence, there does not exist an optimal $\Delta \hat{\mathbf{p}} \in \Omega^{\S}_p$. $\hfill \blacksquare$

\vspace{-10pt}
\subsection{Proof of Closed-form Expression of $D_{u,\text{upper}}$}
\label{appendix_amplitude_bound}

To find $D_{\text{upper}}(l_u,\mathbf{l}_{\phi})$, we can find the upper bound of each component, i.e., $D_{u,\text{upper}}(l_u)$ as an upper bound of minimum $D_u(\mathbf{u}_a,\mathbf{u}_b)$ given ${\rm{card}}\{ \Omega_u \} = 2^{l_u}$, and $D_{p,\text{upper}} ( \mathbf{l}_{\phi} ) $ as the upper bound of minimum $D_p(\mathbf{u})$ given $\mathbf{l}_{\phi}$. Since the MCD is the minimum value among $D_u(\mathbf{u}_a,\mathbf{u}_b)$ and $D_p(\mathbf{u})$, $D_{\text{upper}}(l_u,\mathbf{l}_{\phi})$ can be given by:
\begin{align}\label{eq:bound2}
    D_{\text{upper}}(l_u,\mathbf{l}_{\phi}) =\min \left[ D_{u,\text{upper}}(l_u), D_{p,\text{upper}}(\mathbf{l}_{\phi}) \right].
\end{align}
Let us $l_{\phi,\text{max}}$ be the highest value of $\mathbf{l}_{\phi}$, then $D_{p,\text{upper}} ( \mathbf{l}_{\phi} ) $ can be easily obtained from (\ref{eqn:closed_form_phase_distance}), (\ref{eqn:first_phase_distance}), and (\ref{eqn:second_phase_distance}) as follows:
\begin{align}
    D_{p,\text{upper}} ( \mathbf{l}_{\phi} ) = \sin \left( \frac{\pi}{2^{l_{\phi,\text{max}}}} \right).
\end{align}
Since amplitude vectors are real positive values, finding $D_{u,\text{upper}}(l_u)$ is equivalent to packing $2^{l_u}$ metric balls into a surface of positive-valued $K$-dimensional unit sphere with an area of $S_K = 2(\frac{\pi}{4})^{\frac{K}{2}} / \Gamma(\frac{K}{2})$. Following \cite{mukkavilli2003beamforming}, the volume of a metric ball with chordal distance $D_u$, i.e., $S(D_u)$, is equal to the area of a spherical cap whose chordal distance between its center and its boundary is $D_u$. It is easy to verify that the colatitude angle $\Phi$ of that spherical cap is $D_u = \sin(\Phi)$. Then, the volume of the metric ball is given by \cite{li2011concise}:
\begin{align}
    S(D_u) = \frac{ \pi^{\frac{K-1}{2}} \mathcal{B}\left(\sin^2(\Phi),\frac{K-1}{2},\frac{1}{2}\right) } { \Gamma(\frac{K-1}{2}) } = \frac{ \pi^{\frac{K-1}{2}} \mathcal{B}(D_u^2,\frac{K-1}{2},\frac{1}{2}) } { \Gamma(\frac{K-1}{2}) },
\end{align}
where $\mathcal{B}(x,y,z)$ is the incomplete Beta function. For sufficiently small $D_u^2$, it is easy to prove that $\mathcal{B}\left(D_u^2,\frac{K-1}{2},\frac{1}{2}\right) \approx {2 D_u^{K-1}} / {(K-1)} $. Thus, the Hamming bound is given by \cite{dai2008quantization}:
\begin{align}
    2^{l_u} \leq \frac{S_K}{S( \frac{D_{u}}{2} )} \approx 
    \frac{\pi^{\frac{1}{2}} \Gamma(\frac{K+1}{2})}{\Gamma(\frac{K}{2})D^{K-1}_{u}}.
\end{align}
Thus, $D_{u,\text{upper}}(l_u)$ can be given as:
\begin{align}
    D_{u,\text{upper}}(l_u) = \left( \frac{\pi^{\frac{1}{2}} \Gamma(\frac{K+1}{2})}{\Gamma(\frac{K}{2}) } \right)^{\frac{1}{K-1}} 2^{-\frac{l_u}{K-1}}.
\end{align}
\vspace{-10pt}
\subsection{Proof of the probability density function in (\ref{eqn:simplified_density_func2})}

\label{subsection:proof_PDF}

The term $\text{tr}\left( \bar{\mathbf{Y}}_d^H \bar{\mathbf{Y}}_d \ \bar{\mathbf{v}}_d^* \bar{\mathbf{v}}_d^T \right)$ in (\ref{eqn:simplified_density_func2}) can be formulated as a function of $\phi_d$ and given by:
\begin{align}
    & \text{tr}\left( \bar{\mathbf{Y}}_d^H \bar{\mathbf{Y}}_d \bar{\mathbf{v}}_d^* \bar{\mathbf{v}}_d^T \right) = \sum_{k_1,k_2 \in \bar{\mathcal{D}}} \mathbf{y}^H_{k_1} \mathbf{y}_{k_2} {v}_{k_1} {v}^{*}_{k_2} + \sum_{k_1 \in \bar{\mathcal{D}}} \mathbf{y}^H_{k_1} \mathbf{y}_{d} {v}_{k_1} {v}^{*}_{d} \nonumber \\
    &  + \sum_{k_2 \in \bar{\mathcal{D}}} \mathbf{y}^H_{d} \mathbf{y}_{k_2} {v}_{d} {v}^{*}_{k_2} + \mathbf{y}^H_{d} \mathbf{y}_{d} {v}_{d} {v}^{*}_{d}.
    \label{eqn:quadratic_part}
\end{align}
The first and fourth terms in (\ref{eqn:quadratic_part}) are independent of $\phi_d$. Thus, we reformulate the second and third terms of (\ref{eqn:quadratic_part}) as follows:
{\small{
\begin{align}
    \sum_{k_1 \in \bar{\mathbf{D}}} \mathbf{y}^H_{k_1} \mathbf{y}_{d} {v}_{k_1} {v}^{*}_{d} + \sum_{k_2 \in \bar{\mathcal{D}}} \mathbf{y}^H_{d} \mathbf{y}_{k_2} {v}_{d} {v}^{*}_{k_2} = 2 \text{Re} \left\{ \mu_d e^{-j\phi_d} \right\},
\end{align}
}}
where $\mu_d$ is given by:
\begin{align}
    \mu_d = \sum_{k \in \bar{\mathcal{D}}} \mathbf{y}^H_{k} \mathbf{y}_{d} \tilde{u}_{k} \tilde{u}_{d} e^{j\bar{\phi}_k} = \sum_{{k} \in \bar{\mathcal{D}}} z^{(u)}_{{k},d} e^{j\bar{\phi}_{{k}}}.
\end{align}
{The remaining terms in (\ref{eqn:simplified_density_func2}) are independent of $\phi_d$, thus can be grouped into two constant values $\bar{c}_1$ and $\bar{c}_2$ given by:
\begin{align}
    & \bar{c}_1 = \frac{\exp\left(-\frac{\text{tr}(\bar{\mathbf{Y}}_d^H\bar{\mathbf{Y}}_d)}{\sigma^2} + \frac{ \sum_{k_1,k_2 \in \mathcal{D}} \mathbf{y}^H_{k_1} \mathbf{y}_{k_2} {v}_{k_1} {v}^{*}_{k_2} + \mathbf{y}^H_{d} \mathbf{y}_{d} {v}_{d} {v}^{*}_{d} }{\sigma^2(\sigma^2 + \Vert \bar{\mathbf{v}}_d \Vert^2)} \right)}{\pi^{(n+1)M}\left[ (\sigma^2 + \Vert \bar{\mathbf{v}}_d \Vert^2 )\sigma^{2n} \right]^{M}}, \\
    & \bar{c}_2 = \frac{2}{\sigma^2(\sigma^2 + \Vert \bar{\mathbf{v}}_d \Vert^2 )}.
\end{align}
Finally, the PDF in (\ref{eqn:density_func2}) can be simplified as follows:
\begin{align}
    f(\bar{\mathbf{Y}}_d|\bar{\mathbf{v}}_d) = \bar{c}_{1} \exp\left(\bar{c}_{2} \text{Re}\left\{ \mu_d e^{-j{\phi}_{d}} \right\}\right).
\end{align}
}

\end{appendix}
\bibliographystyle{IEEEtran}
{\footnotesize\bibliography{IEEEabrv, references}}

\end{document}